\documentclass[12pt, a4paper]{article}
\usepackage[utf8]{inputenc}
\usepackage{hyperref}
\usepackage{geometry}
\usepackage{graphicx}
\usepackage{subcaption}
\usepackage[numbers]{natbib}
\usepackage{amssymb}                                                                                                                                                                                                                                                                                                                                                                                                                                                                                                                                                                                                                                                                                                                                                                                                                                                                                                                                                                                                                                                                                                                                                                                                                                                                                                                                                                                                                                                                                                                                                                                                                                                                                                                                                                                                                                                                                                                                                                                                                                                                                                                                                                                                                                                                                                                                                                                                                                                                                                                                                                                                                                                                                                                                                                                                                                                                                                                                                                                                                                                                                                                                                                                                                                                                                                                                                                                                                                                                                                                                                                                                                                                                                                                                                                                                                                                                                                                                                                                                                                                                                                                                                                                                                                          
\usepackage{amsmath}
\usepackage{slashed}
\usepackage{anyfontsize}
\usepackage{bm}
\usepackage{slashed}
\usepackage{multirow}
\usepackage{float}
\usepackage[font=small]{caption}
\usepackage{listings}
\newgeometry{lmargin=1.1in, rmargin=1.1in, tmargin=1in, bmargin=1in}
\title{The effective Dirac algebra by gauge field interaction in relativistic electrodynamics}
\author{B.T.T.Wong\footnote{CERN, u3500478@connect.hku.hk}}
\date{}

\begin{document}

\maketitle
\begin{abstract}
Conventional relativistic electrodynamics is set on flat Minkowski spacetime, where all computable quantities are calculated from the flat metric $\eta_{\mu\nu}$. We can redefine the metric of spacetime from the Dirac algebra. In this paper, we study how an electrodynamic interaction can alter the normal gamma matrix to an effective one and result in a shift in the metric perturbatively. The curvature properties inferred from the curved metric are also investigated. We also study how the spin operator is changed under the interaction that contribute to an effective spin operator and how the spin of an electron will be slightly deviated from $1/2$. Then we perform canonical quantization of the effective Dirac algebra. Finally we apply our results to the relativistic hydrogen case and demonstrate how such system curves the spacetime metric.
\end{abstract}

\section{Introduction}
The reconciliation of special relativity and quantum mechanics has been achieved nicely by Dirac in the 1920s which laid the foundation of relativistic quantum mechanics. The physics works perfectly well in the flat spacetime, i.e. Minkowski spacetime described by the flat metric tensor $\eta_{\mu\nu}$. However, the unification of general relativity in general metric $g_{\mu\nu}$ and quantum mechanics remains one of the hardest problems as quantization of gravity is non-renormalizable, which means that the infinities aroused in the theory cannot be absorbed by a finite number of counter terms, and hence such theory is not predictive\cite{ref1,ref2,ref3,ref4,ref5,ref6}. It will be an essential issue to see, if possible, any connection between the Dirac theory and Einstein's general relativity in both classical case and quantum case. Previous studies on metric modification by electromagnetic potential in line with Einstein's equivalent principle have been performed in references \cite{ext1,ext2}. And there are numerous studies on effective spacetime geometry, for example in references \cite{ext3,ext4,ext5,ext6,ext7,ext8,ext9},  which treat general relativity as an effective field theory. In this paper, we will take the approach of effective Dirac $\gamma^{\mu}$ matrices by electromagnetic potential, and redefine the metric tensor in terms of effective gamma matrices. This will give a curved metric $g_{\mu\nu}$ that is governed by gauge field interactions.

The Dirac equation is the foundation of relativistic electrodynamics which explains the origin of anti-particles. The solution of the Dirac equation represents the relativistic spin-$\frac{1}{2}$ fermions and anti-fermions \cite{DiracAntiMatter, Peskin}. The Dirac equation can be obtained by minimizing the classical Dirac action (in natural units) \cite{Dirac, Dirac3, Peskin},
\begin{equation}
S = \int d^4 x \bigg(-\frac{1}{4}F_{\mu\nu}F^{\mu\nu} + i \bar{\psi} \gamma^\mu \partial_{\mu}\psi - m \bar{\psi}\psi \bigg) \,,
\end{equation}
where $F_{\mu\nu}=\partial_{\mu}A_{\nu}-\partial_{\nu}A_{\mu}$ is the gauge field strength, $\psi$ is the Dirac spinor, $\gamma^{\mu}$ is the Dirac gamma matrix and $m$ is the mass of the fermion. The equation of motion for fermion would be the Dirac equation $i\gamma^\mu \partial_{\mu} \psi -m \psi = 0$ \cite{Dirac}. When the interaction is turned on, the action is modified by a minimum substitution of covariant derivative $\partial_{\mu} \rightarrow D_{\mu} = \partial_{\mu}+ieA_{\mu}$ , where $A_{\mu}$ is the interaction spin-1 photon gauge field \cite{Dirac, Dirac3, Peskin},
 \begin{equation}
S = \int d^4 x \bigg(-\frac{1}{4}F_{\mu\nu}F^{\mu\nu} + i \bar{\psi} \gamma^\mu D_{\mu}\psi - m \bar{\psi}\psi \bigg) \,.
\end{equation}
By minimizing the action the Dirac equation is modified to
\begin{equation} \label{eq:DiracEq}
i \gamma^{\mu} \partial_{\mu} \psi -e\gamma^{\mu}A_{\mu}\psi - m \psi = 0 \,.
\end{equation}

Classical field theory is set on flat Minkowski spacetime and the gamma matrices satisfy the Dirac algebra, which is \cite{Dirac}
\begin{equation} \label{eq:DiracAlgebra}
\{ \gamma^{\mu}, \gamma^{\nu} \} = 2 \eta^{\mu\nu} \pmb{1} \,,
\end{equation}
where $\pmb{1}$ is the $4\times 4$ identity matrix. Therefore the anti-commutation relation of two gamma matrices just gives the natural Minkowski metric tensor. We can look into the opposite perspective by defining the metric of the field theory from its form of gamma matrices.

 In this paper we would like to study on how gauged interaction can modify the current flat spacetime by means of absorbing the interaction in the $\gamma^{\mu}$ vertex in such a way that we redefine the metric using the new effective $\gamma^{\mu}$ vertex $\Gamma^{\mu}$. The metric is then defined by
\begin{equation}
\{ \Gamma^{\mu}, \Gamma^{\nu} \} = 2 g^{\mu\nu} \pmb{1} \,.
\end{equation} 
Perturbatively,
\begin{equation}
\{ \Gamma^{\mu}, \Gamma^{\nu} \} = 2 (\eta^{\mu\nu} -h^{\mu\nu}) \pmb{1} \,,
\end{equation} 
where $h^{\mu\nu } =h^{\mu\nu }(e, A^{\mu} , A^{\nu}) $ is expected to be dependent on the charge coupling and electromagnetic potential. In such way the spacetime is curved by the existence of the electromagnetic interaction. Although such curvature might be small or even negligible, it is worth to study how the new metric behaves and how the spacetime geometry is changed. 

Next we can find out the effective $\gamma^5$ matrix and effective spin operator. The spin of the fermion will be expected to change and deviated from $\frac{1}{2}$, very slightly, due to the effect of gauge field interaction.

The above ideas are in analogy to that, with the electromagnetic interaction and upon renormalization, the electron $g$-factor of the magnetic moment experiences a small shift \cite{Julian},
\begin{equation}
    g = 2 \bigg( 1 + \frac{\alpha}{2\pi} + O(\alpha^2) \bigg) \,,
\end{equation}
where $\alpha = e^2 /4\pi$ is the fine-structural constant in natural unit. But here we study the geometry sense in the classical case, such that we have a very small correction by $h^{\mu\nu}$, while for the spin we would have a small correction for $1/2$.

\section{Effective Dirac Algebra }
\label{sec:1}
We would revisit the Dirac equation and investigate it in a new perspective. For our purpose we will keep the constants $\hbar$ and $c$ in the equations. 
We rewrite the free Dirac equation in the form of
\begin{equation} \label{eq:1}
\psi(x) = \frac{i\hbar}{mc} \gamma^{\mu}\partial_{\mu}\psi(x) = \frac{i\lambda}{2\pi}\gamma^{\mu}\partial_{\mu}\psi(x)\,,
\end{equation}
where $\lambda = \frac{h}{mc}$ is the Compton wavelength. For the Dirac equation with electromagnetic interaction, we rewrite equation (\ref{eq:1}) as
\begin{equation} \label{eq:2}
\psi(x) = \frac{i\hbar}{mc} \frac{1}{1+ \frac{he}{mc} \gamma^{\rho}A_{\rho}(x)} \gamma^{\mu} \partial_{\mu} \psi(x) =\frac{i\lambda}{2\pi}\, \frac{1}{1+ \frac{\lambda e}{2\pi}\gamma^{\rho}A_{\rho}(x)} \gamma^{\mu}\partial_{\mu}\psi(x) \,.
\end{equation}
Or in tensor notation we have
\begin{equation}
\psi_c =\frac{i\lambda}{2\pi}\left( \frac{1}{1+ \frac{\lambda e}{2\pi}\gamma^\rho A_\rho} \right)_{ca} \gamma^{\mu}_{ab} \partial_{\mu}\psi_b \,.
\end{equation}

Therefore, with interaction, we have an extra term contributed by the photon gauge field $A_{\mu}$. Using the Feynman slash notation $\slashed{A}= \gamma^{\mu}A_{\mu}$, define the functional in equation (\ref{eq:2}) as
\begin{equation} \label{eq:functional}
f[\slashed{A}(x)] = \frac{1}{1+ \frac{\lambda e}{2\pi}\slashed{A}(x)} \,.
\end{equation}
The functional can be expanded perturbatively,
\begin{equation} \label{eq:expansion}
\begin{aligned}
f[\slashed{A}(x)] &= \frac{1- \frac{\lambda e}{2\pi} \slashed{A}}{1-\frac{\lambda^2 e^2}{4\pi^2} A^2} \\
&= \bigg( 1+ \frac{\lambda^2 e^2}{4\pi^2} A^2 + \frac{\lambda^4 e^4}{16\pi^2}A^4 + O(e^6)  \bigg) \big( 1- \frac{\lambda e}{2\pi} \slashed{A} \big) \\
&= 1-\frac{\lambda e}{2\pi} \slashed{A} + \frac{\lambda^2 e^2}{4\pi^2}A^2 -  \frac{\lambda^3 e^3}{8\pi^2}\slashed{A}A^2 +  \frac{\lambda^4 e^4}{16\pi^2}A^4 + O(e^5) \,.
\end{aligned}
\end{equation}
Thus the series can be decomposed into the sum of odd and even powers of $\slashed{A}$, where for odd-power series it is $\gamma^{\rho}$ dependent, while of even-power series it is independent of $\gamma^{\rho}$. Define
\begin{equation} \label{eq:EvenOddf}
f_{\rm{even}}[ A_{\rho}  ] = \sum_{k=0,2,4,\cdots}^{\infty} \bigg(\frac{\lambda e}{2\pi}\bigg)^k A^{k} = \frac{1}{1- \frac{\lambda^2 e^2}{4\pi^2} A^2}  \,,
\end{equation}
\begin{equation}
f_{\rm{odd}}[\gamma^{\rho} , A_{\rho}  ] = \sum_{k=1,3,5,\cdots}^{\infty} \bigg(\frac{\lambda e}{2\pi}\bigg)^k \gamma_{\rho}A^{\rho}A^{k-1}= \frac{\lambda e}{2\pi} \slashed{A} f_{\rm{even}}\,,
\end{equation} 
and we have
\begin{equation}
f[\slashed{A}] = f_{\rm{even}}[ A_{\rho}  ] - f_{\rm{odd}}[\gamma^{\rho} , A_{\rho}  ]\,.
\end{equation}

Using $c=1$ in natural unit, we can define the effective mass $\mathcal{M}(x)$ in \ref{eq:2} in terms of the free mass $m$,
\begin{equation}
\mathcal{M}(x) = m\bigg( 1 + \frac{\lambda e}{2 \pi}\gamma^{\mu}A_{\mu}(x) \bigg) \,.
\end{equation}
Alternatively, in equation (\ref{eq:2}), we can define the effective gamma matrix as
\begin{equation}
\Gamma^{\mu}_L = f[\slashed{A}] \gamma^{\mu} \,.
\end{equation}
The subscript $L$ indicates that $f[\slashed{A}]$ is on the left-hand side. Yet we can also define the effective gamma matrix by having $f[\slashed{A}]$ on the right hand side, but we need to introduce a commutator for compensating such change
\begin{equation}
\psi = \frac{i\lambda}{2\pi} \gamma^{\mu} f[\slashed{A}] \partial_{\mu}\psi  - \frac{i\lambda}{2\pi} [\gamma^{\mu} , f[\slashed{A}]] \partial_{\mu}\psi \,.
\end{equation}
Then we remain to evaluate the commutator. Using the identity $\{\gamma^{\mu} , \slashed{A} \} = 2A^{\mu}$, one can show that
\begin{equation}
f[\slashed{A}] \gamma^{\mu} = \Big(\gamma^{\mu} - \frac{\lambda e}{\pi} A^{\mu} \Big) f_{\rm{even}} + \gamma^{\mu}f_{\rm{odd}} \,.
\end{equation}
It follows that
\begin{equation}
[\gamma^{\mu} , f[\slashed{A}]] = \frac{\lambda e}{\pi} A^{\mu}f_{\rm{even}} -2\gamma^{\mu}f_{\rm{odd}} = \frac{\lambda e}{\pi} ( A^{\mu} - \gamma^{\mu} \slashed{A}  ) f_{\rm{even}} \,.
\end{equation}
Using equation (\ref{eq:EvenOddf}), then we finally obtain
\begin{equation}
[\gamma^{\mu} , f[\slashed{A}]] = \frac{\lambda e}{\pi} \bigg(\frac{A^{\mu} - \gamma^{\mu} \slashed
A}{1-\frac{\lambda^2 e^2}{4\pi^2}A^2}  \bigg) \,.
\end{equation}
Therefore the Dirac equation with electromagnetic interaction can be written as order of $e$ as
\begin{equation}
\begin{aligned}
\psi &= \frac{i\lambda}{2\pi} \Gamma^{\mu}_R \partial_{\mu}\psi  - \frac{i\lambda^2 e}{2\pi^2} \bigg(\frac{A^{\mu} - \gamma^{\mu} \slashed
A}{1-\frac{\lambda^2 e^2}{4\pi^2}A^2}  \bigg) \partial_{\mu}\psi \\
&= \frac{i\lambda}{2\pi} \Gamma^{\mu}_R \partial_{\mu}\psi + \bigg( \frac{\lambda^2 e}{2\pi^2} + \frac{\lambda^4 e^3}{8\pi^4} + O(e^5)\bigg) ( A^{\mu} - \gamma^{\mu} \slashed{A} ) i \partial_{\mu}\psi \,.
\end{aligned}
\end{equation}

Finally, we would like to study how can the interaction change the geometry of flat spacetime. First we know that classical field theory is defined in flat space time by the flat metric. Now with the interaction, the effective gamma matrix $\Gamma_L^{\mu}$ follows the new Dirac algebra, 
\begin{equation}
\{ \Gamma^{\mu}_L , \Gamma^{\nu}_L \} = 2g^{\mu\nu} \pmb{1}  \,,
\end{equation}
where $g^{\mu\nu}$ is the curved metric, and it depends on $\lambda, e , \gamma^{\rho}$ and the gauge field $A^{\rho}(x)$. The full derivation is given as follow. First consider
\begin{equation} \label{eq:Eq1}
\{ \Gamma^{\mu}_L , \Gamma^{\nu}_L \} = \{ f[\slashed{A}] \gamma^{\mu} , f[\slashed{A}]\gamma^\nu  \} = f[\slashed{A}]\gamma^{\mu}f[\slashed{A}] \gamma^{\nu} + f[\slashed{A}] \gamma^{\nu} f[\slashed{A}] \gamma^{\mu} \,.
\end{equation}
Then using the fact that
\begin{equation} \label{eq:Eq2}
[ \gamma^{\mu} , f[\slashed{A}] \, ] = \gamma^{\mu}f[\slashed{A}] - f[\slashed{A}]\gamma^{\mu} 
\implies \gamma^{\mu} f[ \slashed{A}] \, [ \gamma^{\mu} , f[\slashed{A}] \, ] + f[\slashed{A}]\gamma^{\mu} 
\end{equation}
and substituting equation (\ref{eq:Eq2}) into equation (\ref{eq:Eq1}), we obtain
\begin{equation} \label{eq:Eq3}
\begin{aligned}
&\quad \{ \Gamma^{\mu}_L , \Gamma^{\nu}_L \}\\
&= f[\slashed{A}] \, \bigg\{ [ \gamma^{\mu} , f[\slashed{A}] \, ]+ f[\slashed{A}]\gamma^{\mu} \bigg\} \gamma^\nu + f[\slashed{A}] \, \bigg\{ [ \gamma^{\nu} , f[\slashed{A}]\, ]+ f[\slashed{A}]\gamma^{\nu} \bigg\} \gamma^\mu  \\
&=f[\slashed{A}] \,[ \gamma^{\mu} , f[\slashed{A}] \,] \gamma^{\nu} +  f[\slashed{A}] \,[ \gamma^{\nu} , f[\slashed{A}]\, ] \gamma^{\mu} + f^2 [\slashed{A}] (\gamma^{\mu}\gamma^{\nu} + \gamma^{\nu}\gamma^{\mu} ) \\
&= 2 f^2 [\slashed{A}] \eta^{\mu\nu} + f[\slashed{A}] \, \frac{\lambda e}{\pi} \bigg(\frac{A^{\mu} - \gamma^{\mu} \slashed{A}}{1-\frac{\lambda^2 e^2}{4\pi^2}A^2}\bigg) \gamma^\nu + f[\slashed{A}]\,\frac{\lambda e}{\pi} \bigg(\frac{A^{\mu} - \gamma^{\mu} \slashed{A}}{1-\frac{\lambda^2 e^2}{4\pi^2}A^2} \gamma^\mu \bigg)    \\
&= 2 f^2 [\slashed{A}] \eta^{\mu\nu}  + \frac{\lambda e}{\pi} f[\slashed{A}] \bigg( \frac{1}{1-\frac{\lambda^2 e^2}{4\pi^2}A^2} \bigg) ( A^{\mu} \gamma^\nu + A^{\nu}\gamma^{\mu} - \gamma^\mu \slashed{A}\gamma^{\nu} - \gamma^{\nu}\slashed{A}\gamma^{\mu} ) \,.
\end{aligned}
\end{equation}
To further simply terms of $\gamma^\mu \slashed{A}\gamma^{\nu} , \gamma^{\nu}\slashed{A}\gamma^{\mu}$, we use the identity
\begin{equation}
\{ \gamma^\mu , \slashed{A} \} = 2A^{\mu} \implies \slashed{A}\gamma^\mu = 2A^\mu - \gamma^\mu\slashed{A} \,.
\end{equation}
Then the last two terms in the last line of equation \ref{eq:Eq3} simplify to
\begin{equation}
\begin{aligned}
&\quad -(\gamma^\mu \slashed{A}\gamma^{\nu} + \gamma^{\nu}\slashed{A}\gamma^{\mu} ) \\
&= \Big( \gamma^\mu (2A^\nu - \gamma^\nu\slashed{A} ) + \gamma^{\nu} (2A^\mu - \gamma^\mu\slashed{A} )  \Big) \\
&= - \Big( 2\gamma^\mu A^\nu + 2 \gamma^\nu A^\mu -(\gamma^\mu \gamma^\nu + \gamma^\nu \gamma^\mu )\slashed{A} \Big) \\
&= -(  2\gamma^\mu A^\nu + 2 \gamma^\nu A^\mu -2 \eta^{\mu\nu} \slashed{A}  ) \,.
\end{aligned}
\end{equation}
Substituting this result back into equation(\ref{eq:Eq3}), we finally obtain the desired equation for the anti-commutator for the effective Dirac algebra,
\begin{equation} \label{eq:EffectiveDiracAlgebra}
\{ \Gamma^{\mu}_L , \Gamma^{\nu}_L \} = 2 f^2 [\slashed{A}] \eta^{\mu\nu} + \frac{\lambda e}{\pi} f[\slashed{A}] \bigg( \frac{1}{1-\frac{\lambda^2 e^2}{4\pi^2}A^2} \bigg) ( 2\eta^{\mu\nu}\slashed{A}-\gamma^\mu A^\nu - \gamma^\nu A^\mu )\,.
\end{equation}
Therefore we have the full form of curved (inverse) metric matrix as follow,
\begin{equation} \label{eq:4}
g^{\mu\nu} \pmb{1} = f^2 [\slashed{A}] \eta^{\mu\nu} + \frac{\lambda e}{2\pi} f[\slashed{A}] \bigg( \frac{1}{1-\frac{\lambda^2 e^2}{4\pi^2}A^2} \bigg) ( 2\eta^{\mu\nu}\slashed{A}-\gamma^\mu A^\nu - \gamma^\nu A^\mu ) \,.
\end{equation}
The metric can be obtained by taking the trace both sides of \ref{eq:4},
\begin{equation} \label{eq:metricc}
g_{\mu\nu} = \frac{1}{4}\mathrm{Tr}\, \bigg\{ f^2 [\slashed{A}] \eta_{\mu\nu} + \frac{\lambda e}{2\pi} f[\slashed{A}] \bigg( \frac{1}{1-\frac{\lambda^2 e^2}{4\pi^2}A^2} \bigg) ( 2\eta_{\mu\nu}\slashed{A}-\gamma_\mu A_\nu - \gamma_\nu A_\mu )\bigg\} \,.
\end{equation}

Next we are interested in how this can be resulted in a weak perturbation theory of the metric.
Since $f[\slashed{A}]$ can be worked out perturbatively, $g_{\mu\nu}$  can be expected to be a expansion series of the coupling $e, \lambda$ and gauge field $A^{\rho}$,
\begin{equation}
g^{\mu\nu}( \lambda , e,  A^{\rho} ) = \eta^{\mu\nu} - h^{\mu\nu}  (\lambda , e, A^{\rho} )\,.
\end{equation}
Therefore, the effective Dirac algebra can be regarded as the original Dirac algebra with some perturbative corrections from the interaction,
\begin{equation}
\{ \Gamma^{\mu}_L , \Gamma^{\nu}_L \} = \{ \gamma^{\mu} , \gamma^{\nu} \} + {\rm perturbative\,\,corrections} \,.
\end{equation}
Here we would like to carry out perturbation theory so we expand in the powers of the coupling $O(e^n)$, and do it up to the second order. Consider
\begin{equation}
f[\slashed{A}] = \frac{1}{1+ \frac{\lambda e}{2\pi}\slashed{A}} = 1- \frac{\lambda e}{2\pi}\slashed{A} + \frac{\lambda^2 e^2}{4\pi^2}A^2 + O(e^3) \,.
\end{equation}
Thus
\begin{equation}
\begin{aligned}
f^2 [\slashed{A}] &=  \bigg( 1- \frac{\lambda e}{2\pi}\slashed{A} + \frac{\lambda^2 e^2}{4\pi^2}A^2 + O(e^3) \bigg) \bigg( 1- \frac{\lambda e}{2\pi}\slashed{A} + \frac{\lambda^2 e^2}{4\pi^2}A^2 + O(e^3) \bigg) \\
&= 1- \frac{\lambda e}{2\pi}\slashed{A} + \frac{\lambda^2 e^2}{4\pi^2}A^2 -\frac{\lambda e}{2\pi}\slashed{A} + \frac{\lambda^2 e^2}{4\pi^2}\slashed{A}^2 + \frac{\lambda^2 e^2}{4\pi^2}A^2 + O(e^3) \\
&= 1- \frac{\lambda e}{\pi}\slashed{A} + \frac{3\lambda^2 e^2}{4\pi^2}A^2 + O(e^3) \,.
\end{aligned}
\end{equation}
Therefore the first term in equation (\ref{eq:EffectiveDiracAlgebra}) reads
\begin{equation} \label{eq:resultFirstterm}
2 f^2 [\slashed{A}]  = 2\eta^{\mu\nu} + \bigg( \frac{2\lambda e}{\pi}\slashed{A} + \frac{3\lambda^2 e^2}{2\pi^2}A^2 + O(e^3) \bigg) \eta^{\mu\nu} \,.
\end{equation}
Next we consider the expansion for the second terms in equation(\ref{eq:EffectiveDiracAlgebra}). Notice that
\begin{equation} \label{eq:5}
\frac{1}{1-\frac{\lambda^2 e^2}{4\pi^2}A^2} = 1 + \frac{\lambda^2 e^2}{4\pi^2}A^2 + O(e^4) \,,
\end{equation}
then all together for equations (\ref{eq:EffectiveDiracAlgebra}), (\ref{eq:resultFirstterm}) and (\ref{eq:5}), we have
\begin{equation} \label{eq:resultSecondterm}
\begin{aligned}
&\quad \frac{\lambda e}{\pi} \bigg( 1- \frac{\lambda e}{2\pi}\slashed{A} + \frac{\lambda^2 e^2}{4\pi^2}A^2 + O(e^3) \bigg) \bigg( 1 + \frac{\lambda^2 e^2}{4\pi^2}A^2 O(e^4)\bigg)( 2\eta^{\mu\nu}\slashed{A}-\gamma^\mu A^\nu - \gamma^\nu A^\mu )\\
&= \frac{\lambda e}{\pi} \bigg( 1-\frac{\lambda e}{2\pi}\slashed{A} + \frac{3\lambda^2 e^2}{4\pi^2}A^2 + O(e^3)  \bigg) ( 2\eta^{\mu\nu}\slashed{A}-\gamma^\mu A^\nu - \gamma^\nu A^\mu ) \\
&= \bigg(\frac{2\lambda e}{\pi}\slashed{A} - \frac{\lambda^2 e^2}{\pi^2}A^2  \bigg) \eta^{\mu\nu} - \bigg( \frac{\lambda e}{\pi} - \frac{\lambda^2 e^2}{2\pi^2}\slashed{A} \bigg) (\gamma^{\mu}A^{\nu} + \gamma^{\nu}A^{\mu}) + O(e^3) \,.
\end{aligned}
\end{equation}
Combining the results in equation(\ref{eq:resultFirstterm}) and equation(\ref{eq:resultSecondterm}), we obtain
\begin{equation} \label{eq:SecondOrderEffectiveDiracAlgebra}
\{ \Gamma^{\mu}_L , \Gamma^{\nu}_L \} = 2g^{\mu\nu} \pmb{1} = \bigg(2 + \frac{\lambda^2 e^2}{2\pi^2}A^2 \bigg)\eta^{\mu\nu}\pmb{1} - \bigg( \frac{\lambda e}{\pi} - \frac{\lambda^2 e^2}{2 \pi^2} \slashed{A}   \bigg) B^{\mu\nu} + O(e^3) \,,
\end{equation}
where we define the symmetric tensor $B^{\mu\nu} =\gamma^{\mu}A^{\nu} + \gamma^{\nu}A^{\mu}$.

The inverse perturbation metric matrix is thus
\begin{equation}
h^{\mu\nu} \pmb{1} = -\frac{\lambda^2 e^2}{4\pi^2} A^2\eta^{\mu\nu} + \bigg(\frac{\lambda e}{2\pi} - \frac{\lambda^2 e^2}{4 \pi^2} \slashed{A}   \bigg) B^{\mu\nu} + O(e^3) \,.
\end{equation}
To obtain the metric tensor, we take the trace both sides, thus the full metric is just
\begin{equation}
ds^2 = \frac{1}{8}{\mathrm{Tr}} \{ \Gamma_{L\,\mu} , \Gamma_{L\,\nu} \} dx^\mu dx^\nu \,.
\end{equation}
Notice that ${\mathrm{Tr}}\gamma^\mu =0$ and $\mathrm{Tr}\gamma^\mu \gamma^\nu = 4\eta^{\mu\nu}$, then
\begin{equation}
\begin{aligned}
4g_{\mu\nu} &= \bigg(1 + \frac{\lambda^2 e^2}{4\pi^2}A^2 \bigg)\eta^{\mu\nu} {\mathrm{Tr}\,\pmb{1}} + \frac{\lambda^2 e^2}{4\pi^2} \Big( A_{\nu}A^{\rho} {\mathrm{Tr}}(\gamma_\rho \gamma_\mu) + A_\mu A_{\rho} {\mathrm{Tr}} (\gamma_\rho \gamma_\nu )  \Big) \\
&= 4\bigg(1 + \frac{\lambda^2 e^2}{4\pi^2}A^2 \bigg)\eta_{\mu\nu} + \frac{2\lambda^2 e^2}{\pi^2}A_{\mu} A_{\nu} \,.
\end{aligned}
\end{equation}
Hence, we obtain 
\begin{equation} \label{eq:metriccc}
g_{\mu\nu} = \bigg(1 + \frac{\lambda^2 e^2}{4\pi^2}A^2 \bigg)\eta_{\mu\nu} +  \frac{\lambda^2 e^2}{2\pi^2}A_{\mu} A_{\nu} \,.
\end{equation}
Therefore, the perturbation metric tensor up to second order of coupling is
\begin{equation} \label{eq:h}
h_{\mu\nu} = \frac{\lambda^2 e^2}{4\pi^2} ( A^2 \eta_{\mu\nu} + 2A_{\mu} A_{\nu}  ) \,.
\end{equation}
Therefore, the spacetime geometry is changed by the tiny amount given by equation (\ref{eq:h}).

\section{Curvature and geometric properties from effective Dirac algebra}
In this section, we would like to calculate how the geometry changes due to metric change arised from the effective Dirac algebra. Since the metric is perturbated, we can use the linearized perturbation theory of general relativity.

The Christoffel connection is given by \cite{Gravitation}
\begin{equation}
\mathrm{\Gamma}^{\rho}_{\mu\nu} = \frac{1}{2}\eta^{\rho\lambda}(\partial_{\mu}h_{\nu\lambda} + \partial_{\nu}h_{\lambda\mu} - \partial_{\lambda}h_{\mu\nu}   ) \,.    
\end{equation}
Upon explicit computation,
\begin{equation}
\mathrm{\Gamma}^{\rho}_{\mu\nu} =\frac{\lambda^2 e^2}{8\pi^2}[( \delta^{\rho}_{\nu}\partial_{\mu} + \delta^\rho_{\mu}\partial_{\nu} -\eta_{\mu\nu}\partial^\rho)A^2 +2A_\nu F_{\mu}^{\,\,\,\rho} + 2A_{\mu}F_{\nu}^{\,\,\,\rho} + 2A^{\rho} C_{\mu\nu}  ] \,,
\end{equation}
where $C_{\mu\nu} = \partial_{\mu} A_{\nu} + \partial_{\nu}A_{\mu}$ is a symmetric tensor, and $F_{\mu}^{\,\,\,\rho} = \eta^{\rho \lambda}F_{\mu\lambda}$ with $F_{\mu\lambda}=\partial_{\mu}A_{\lambda}-\partial_{\lambda}A_{\mu} $ the anti-symmetric field strength tensor. Therefore, the geodesic equation is, 
\begin{equation}
\frac{d^2 x^\rho}{dt^2} +   \frac{\lambda^2 e^2}{8\pi^2}[ (\delta^{\rho}_{\nu}\partial_{\mu} + \delta^\rho_{\mu}\partial_{\nu} -\eta_{\mu\nu}\partial^\rho)A^2 +2A_\nu F_{\mu}^{\,\,\,\rho} + 2A_{\mu}F_{\nu}^{\,\,\,\rho} + 2A^{\rho} C_{\mu\nu}  ] \frac{dx^\mu}{dt}\frac{dx^\nu}{dt}=0\,.  
\end{equation}
The Riemannian tensor is given by \cite{Gravitation}
\begin{equation}
R_{\mu\nu\rho\sigma}=\frac{1}{2}( \partial_{\rho}\partial_{\nu} h_{\mu\sigma} +\partial_{\sigma}\partial_{\mu}h_{\nu\rho}-\partial_{\sigma}\partial_{\nu}h_{\mu\rho}-\partial_{\rho}\partial_{\mu}h_{\nu\sigma} ) \,.
\end{equation}
After some lines of algebra, the Riemannian tensor is explicitly computed to be 
\begin{equation}
\begin{aligned}
R_{\mu\nu\rho\sigma} &= \frac{\lambda^2 e^2}{4\pi^2}\bigg(\partial_{\nu}A_{\beta}(\eta_{\mu\sigma}\partial_{\rho}-\eta_{\mu\rho}\partial_{\sigma} )A^{\beta} + \partial_{\mu}A_{\beta}(\eta_{\nu\rho}\partial_{\sigma}-\eta_{\nu\sigma}\partial_{\rho}  )A^\beta \\
& +A_{\alpha}( \eta_{\mu\sigma}\partial_{\rho}-\eta_{\mu\rho}\partial_{\sigma} )\partial_{\nu}A^{\alpha} + A_{\alpha}( \eta_{\nu\rho}\partial_{\sigma} -\eta_{\nu\sigma}\partial_{\rho} )\partial_{\mu}A^{\alpha} \\
&+ \frac{1}{2}\Big((A_{\sigma}\partial_{\rho}-A_{\rho}\partial_{\sigma}  )F_{\nu\mu} +(A_{\mu}\partial_{\nu}-A_{\nu}\partial_{\mu}   )F_{\rho\sigma} +F_{\mu\nu}F_{\sigma\rho} \\
&+\partial_{\rho}A_{\mu}\partial_{\nu}A_{\sigma}-\partial_{\sigma}A_{\mu}\partial_{\nu}A_{\rho} +\partial_{\sigma}A_{\nu}\partial_{\mu}A_{\rho}-\partial_{\rho}A_{\nu}\partial_{\mu}A_{\sigma}\Big)\bigg) \,.
\end{aligned}
\end{equation}
The Ricci tensor is given by \cite{Gravitation}
\begin{equation} \label{eq:6}
R_{\mu\nu} =\frac{1}{2}( \partial_{\sigma}\partial_{\nu}h^{\sigma}_{\,\,\,\mu} + \partial_{\sigma}\partial_{\mu}h^{\sigma}_{\,\,\,\nu}       -\partial_{\mu}\partial_{\nu}h - \Box h_{\mu\nu}  ) \,.
\end{equation}
For simplicity we will impose the Lorentz gauge condition that $\partial_{\mu}A^{\mu} = 0$. The first term of the Ricci tensor in equation (\ref{eq:6}) is
\begin{equation}
\partial_{\sigma}\partial_{\nu}h^{\sigma}_{\,\,\,\mu} =  \frac{\lambda^2 e^2}{4\pi^2} (\partial_{\mu} \partial_{\nu}A^2  + 2\partial_{\nu}A^{\sigma}\partial_{\sigma}A_{\mu} + 2A^{\sigma}\partial_{\sigma}\partial_{\nu}A_{\mu}) \,.
\end{equation}
The second term of the Ricci tensor in equation (\ref{eq:6}) is
\begin{equation}
\partial_{\sigma}\partial_{\mu}h^{\sigma}_{\,\,\,\nu} = \frac{\lambda^2 e^2}{4\pi^2}( \partial_{\nu} \partial_{\mu}A^2  + 2\partial_{\mu}A^{\sigma}\partial_{\sigma}A_{\nu} + 2A^{\sigma}\partial_{\sigma}\partial_{\mu}A_{\nu} ) \,.
\end{equation}
Since $h=h^{\rho}_{\,\,\,\rho} = \frac{\lambda^2 e^2}{4\pi^2}(6A^2)$, the third term in equation (\ref{eq:6}) is just simply $-\partial_{\mu}\partial_{\nu} h= -\frac{3\lambda^2 e^2}{2\pi^2}(\partial_{\mu}\partial_{\nu}A^2) \,.$
Lastly we have the forth term in equation (\ref{eq:6}) as
\begin{equation}
-\Box h_{\mu\nu} =\frac{\lambda^2 e^2}{4\pi^2} \Big(-\eta_{\mu\nu} \Box A^2 - 2( A_{\nu}\Box A_{\mu} + A_{\mu}\Box A_{\nu} + 2\partial^{\alpha}A_{\mu} \partial_{\alpha}A_{\nu}  ) \Big)\,.
\end{equation}
The Ricci scalar is given by \cite{Gravitation}
\begin{equation}
R = \partial_{\mu}\partial_{\nu}h^{\mu\nu} -\Box h =\frac{\lambda^2 e^2}{4\pi^2}(-5\Box A^2 +2\partial_{\alpha}A^\beta \partial_{\beta} A^{\alpha}) \,.
\end{equation}
Finally the Einstein tensor is defined by
\begin{equation}
G_{\mu\nu} = R_{\mu\nu}-\frac{1}{2}\eta_{\mu\nu}R = 8\pi G T_{\mu\nu}\,.
\end{equation}
Thus we obtain
\begin{equation}
\begin{aligned}
G_{\mu\nu}&=\frac{\lambda^2 e^2}{4\pi^2}\Big( 2(\Box\eta_{\mu\nu}-\partial_{\mu}\partial_{\nu} )A^2 + \partial_{\nu}A^\sigma \partial_{\sigma}A_{\mu} + \partial_{\mu}A^\sigma \partial_{\sigma}A_{\nu} + A^{\sigma}\partial_{\sigma}(\partial_{\mu}A_{\nu}+\partial_{\nu}A_{\mu}) \\
&\,\,\,\,-A_{\nu}\Box A_{\mu} -A_{\mu}\Box A_{\nu} -2\partial^{\alpha}A_{\mu}\partial_{\alpha}A_{\nu} -\eta_{\mu\nu}\partial_{\alpha}A^\beta \partial_{\beta}A^\alpha \Big)\,, 
\end{aligned}
\end{equation}
which is a non-linear equation.

\section{Effective $\gamma^5$, effective spin operator and effective spin of the electron}
In this section, we would like to study how the spin of an electron changes under the effective Dirac algebra. First, the $\gamma^5$ is a parity operator defined by $\gamma^5 = i \gamma^0 \gamma^1 \gamma^2 \gamma^3 $, which is the only operator that commutes with all other $\gamma^\mu$ matrices \cite{Peskin}. The spin operator is constructed by the $\gamma^5$ matrix, in the Dirac representation \cite{Peskin},
\begin{equation}
S^i = \frac{1}{2} \gamma^5 \gamma^0 \gamma^i 
\end{equation}
for $i=1,2,3$, where $S^1, S^2 \text{and} S^3$ correspond to the $S_x , S_y , S_z$ spin components respectively. It can be easily found that
\begin{equation}
S^2 = \sum_{i=1}^3 (S^i)^2  =S_x^2 + S_y^2 + S_z^2 = \frac{3}{4}\pmb{1}\,.
\end{equation}
And as $S^2 =s(s+1)\pmb{1}$, this implies $s=\frac{1}{2}$. Thus the spin operator describes the spin-half electron.

We will see how the interaction with gauge field effectively changes the spin of the electron.  The effective $\gamma^5$ operator becomes
\begin{equation}
\Gamma_L^5 = i\Gamma_L^0 \Gamma_L^1 \Gamma_L^2 \Gamma_L^3 = i f[\slashed{A}] \gamma^0 f[\slashed{A}] \gamma^1 f[\slashed{A}] \gamma^2 f[\slashed{A}] \gamma^3 \,.
\end{equation}
Up to second order, after some algebras one finds,
\begin{equation}
\begin{aligned}
\Gamma_L^5 &= \prod_{k=0}^3 \bigg(1-\frac{\lambda e}{2\pi} \slashed{A} +\frac{\lambda^2 e^2}{4\pi^2} A^2   + O(e^3) \bigg) \gamma^k \\
&= i\bigg(  \gamma^0 \gamma^1 - \frac{\lambda^2 e^2}{4\pi^2}\Big(-\gamma^0 \slashed{A} \gamma^1 + 2A^2 \gamma^0 \gamma^1 -\slashed{A}\gamma^0 \slashed{A} \gamma^1   \Big) + O(e^3)  \bigg) \\
&\quad \quad \times \bigg(  \gamma^2 \gamma^3 - \frac{\lambda^2 e^2}{4\pi^2}\Big(-\gamma^2 \slashed{A} \gamma^3 + 2A^2 \gamma^2 \gamma^3 -\slashed{A}\gamma^2 \slashed{A} \gamma^3   \Big) + O(e^3) \bigg) \,.
\end{aligned}
\end{equation}
Then it follows that,
\begin{equation} \label{eq:step1}
\begin{aligned}
\Gamma_L^5 &= \gamma^5+ i\frac{\lambda^2 e^2}{4\pi^2}\bigg(4A^2 \gamma^0\gamma^1 \gamma^2 \gamma^3  -\gamma^0\gamma^1 \gamma^2 \slashed{A}\gamma^3 - \gamma^0 \slashed{A}\gamma^1 \gamma^2 \gamma^3 \\
&\quad\quad\quad\quad\quad\quad\quad\quad - \gamma^0 \gamma^1 \slashed{A} \gamma^2 \slashed{A} \gamma^3 -\slashed{A} \gamma^0 \slashed{A} \gamma^1 \gamma^2 \gamma^3   \bigg) + O(e^3) \,.
\end{aligned}
\end{equation}
With some manipulation, finally we can write $\Gamma_L^5$ as some correction of the original $\gamma^5$ matrix with some extra correction terms 
\begin{equation}
\Gamma_L^5 = \bigg(1 +\frac{3\lambda^2 e^2 }{2\pi^2}A^2 \bigg) \gamma^5 + F(A^{\mu} , \gamma^{\nu} ) + O(e^3) \,, 
\end{equation}
where
\begin{equation}
F(A^{\mu} , \gamma^{\nu} ) = -\frac{i\lambda^2 e^2}{2\pi^2} \bigg( \gamma^0 \gamma^1 \gamma^2 A^{(3)} + A^{(0)}\gamma^1 \gamma^2 \gamma^3 + 2\gamma^0 \gamma^1 A^{(2)}A^{(3)} - \big(\gamma^0 \gamma^1    A^{(2)} + \gamma^1 \gamma^2  A^{(0)} \big) \gamma^3 \slashed{A}  \bigg)\,.
\end{equation}
We can see that only if $A^{\mu}$ vanishes we get back $\gamma^5$. 

Next we compute the effective spin operator. We have
\begin{equation} \label{eq:step2}
\Sigma^i =\frac{1}{2}\Gamma^5_L \Gamma_L^0 \Gamma^i_L \,.
\end{equation}
and we want to express the effective spin operator $\Sigma^i$ in terms of the original $S^i$ with some extra corrections. We can use the result in equation(\ref{eq:step1}),
\begin{equation}
\begin{aligned}
\Sigma^i &= \frac{1}{2}\Bigg(\gamma^5+ i\frac{\lambda^2 e^2}{4\pi^2}\bigg(4A^2 \gamma^0\gamma^1 \gamma^2 \gamma^3  -\gamma^0\gamma^1 \gamma^2 \slashed{A}\gamma^3 - \gamma^0 \slashed{A}\gamma^1 \gamma^2 \gamma^3 \\
&\quad\quad - \gamma^0 \gamma^1 \slashed{A} \gamma^2 \slashed{A} \gamma^3 -\slashed{A} \gamma^0 \slashed{A} \gamma^1 \gamma^2 \gamma^3   \bigg)\Bigg) \times\bigg(\gamma^0 - \frac{\lambda e^2}{4\pi^2}\slashed{A}\gamma^0 +\frac{\lambda^2 e^2}{4\pi^2} A^2 \gamma^0  \bigg) \\ 
&\quad\quad \times \bigg(\gamma^i - \frac{\lambda e^2}{4\pi^2}\slashed{A}\gamma^i +\frac{\lambda^2 e^2}{4\pi^2} A^2 \gamma^i \bigg) \\
&=S^i + \frac{\lambda^2 e^2}{2\pi^2}\Bigg( A^3 \gamma^5 \gamma^0 \gamma^i - \frac
{i}{4}\gamma^0 \gamma^1 \gamma^2 \slashed{A}\gamma^3 \gamma^0 \gamma^i -\frac{i}{4}\gamma^0 \gamma^1 \slashed{A}\gamma^2\slashed{A}\gamma^3\gamma^0\gamma^i \\
&\quad\quad -\frac{i}{4}\gamma^0\slashed{A}\gamma^1 \gamma^2 \gamma^3 \gamma^0 \gamma^i -\frac{i}{4}\slashed{A}\gamma^0 \slashed{A}\gamma^1 \gamma^2 \gamma^3\gamma^0 \gamma^i -\frac{i}{4}\gamma^0 \gamma^1 \gamma^2 \gamma^3 \gamma^0 \slashed{A}\gamma^i \\
&\quad\quad +\frac{1}{2}A^2 \gamma^5\gamma^0\gamma^i -\frac{1}{4}\gamma^5\slashed{A}\gamma^0 \slashed{A}\gamma^i
 \Bigg)
\end{aligned}
\end{equation}
Employing the identity of $\{\gamma^\mu , \slashed{A} \} = 2A^\mu$ and the fact that $\slashed{A} \slashed{A}=A^2$ to the fifth term and the sixth term in the last line of equation(\ref{eq:step2}), we get respectively,
\begin{equation}
-\frac{i}{4}\slashed{A}\gamma^0\slashed{A}\gamma^1 \gamma^2\gamma^3\gamma^0 \gamma^i =
 -\frac{i}{2}A^0 \gamma^1 \gamma^2 \gamma^3 \gamma^0 \gamma^i +\frac{1}{4}A^2 \gamma^5 \gamma^0 \gamma^i
\end{equation}
and
\begin{equation}
-\frac{i}{4}\gamma^0 \gamma^1 \gamma^2 \gamma^3 \gamma^0 \slashed{A} \gamma^i = -\frac{i}{2}\gamma^0 \gamma^1 \gamma^2 \gamma^3 \gamma^0 A^i + \frac{1}{4}\gamma^5 \gamma^0\gamma^i\slashed{A} \,.
\end{equation}
After some simplification, we finally obtain the desired formula,
\begin{equation} \label{eq:10}
\Sigma^i = \bigg( 1+ \frac{3\lambda^2 e^2}{2\pi^2}A^2\bigg)S^i +\frac{\lambda^2 e^2}{4\pi^2} \{S^i, \slashed{A}\} + F(A^{\mu}, \gamma^{\nu} ,\gamma^5) \,,
\end{equation}
where
\begin{equation}
\begin{aligned}
F(\gamma^{\mu} , A^{\nu} ,\gamma^5 ) &=\frac{\lambda^2 e^2}{\pi^2}\bigg(\gamma^5 A^0 \gamma^i + \frac{i}{2} A^i \gamma^1 \gamma^2 \gamma^3 -\frac{i}{4}\gamma^0\gamma^1\gamma^2\slashed{A} \gamma^3 \gamma^0 \gamma^i \\
&\quad\quad\quad\quad -\frac{i}{4}\gamma^0\gamma^1\slashed{A}\gamma^2\slashed{A}\gamma^3\gamma^0\gamma^i -\frac{i}{4}\gamma^5\slashed{A}\gamma^0 \slashed{A} \gamma^i \bigg)
\end{aligned}
\end{equation}
Thus we can see that how the origin spin operator is corrected by the gauge field, coupling and the anti-commutator $\{S^i ,\slashed{A} \}$. We also clearly see that when the gauge field is turned off, we get back the original spin operator $S^i$.

To see how the spin is changed, first for convenience, in equation \ref{eq:10} we define,
\begin{equation}
  G^{i}(S^i, \slashed{A}, A^{\mu} , \gamma^5 ) = \frac{\lambda^2 e^2}{4\pi^2} \{S^i, \slashed{A}\} + F(A^{\mu}, \gamma^{\nu} ,\gamma^5) \,.
\end{equation}
Therefore, let's rewrite equation (\ref{eq:10}) as
\begin{equation}
    \Sigma^i = \bigg( 1+ \frac{3\lambda^2 e^2}{2\pi^2}A^2\bigg)S^i +  G^{i}(S^i, \slashed{A}, A^{\mu} , \gamma^5 ) \,.
\end{equation}
The effective spin operator square $\Sigma^2$ is, therefore,
\begin{equation}
\begin{aligned}
\Sigma^2 = \sum_{i=1}^3 (\Sigma^i)^2 & =\sum_{i=1}^3 \left(  \bigg( 1+ \frac{3\lambda^2 e^2}{2\pi^2}A^2\bigg)S^i +  G^{i}(S^i, \slashed{A}, A^{\mu} , \gamma^5 ) \right)^2 \\
&= \sum_{i=1}^3\left( \bigg( 1+ \frac{3\lambda^2 e^2}{2\pi^2}A^2\bigg)^2(S^i)^2 + \bigg( 1+ \frac{3\lambda^2 e^2}{2\pi^2}A^2\bigg) \{S^i ,G  \} + G^2  \right)\,.
\end{aligned}
\end{equation}
Next for our approximation, we would like to keep the second order terms only, therefore we have
\begin{equation} \label{eq:11}
\begin{aligned}
\Sigma^2 &= \bigg( 1 + \frac{3\lambda^2 e^2}{\pi^2}A^2 \bigg)  \sum_{i=1}^3 (S^i)^2  +\bigg(\sum_{i=1}^3 \{S^i , G\} \bigg) + O(e^4)  \\
&= \frac{3}{4}\pmb{1} + \frac{9\lambda^2 e^2}{4\pi^2}A^2 \pmb{1}  +\bigg(\sum_{i=1}^3 \{S^i , G\} \bigg) + O(e^4) \\
&\equiv \frac{3}{4}\pmb{1} + \Delta \,,
\end{aligned}
\end{equation}
where $\Delta$ is the additional matrix extra terms that
\begin{equation}
\Delta = \frac{9\lambda^2 e^2}{4\pi^2}A^2 \pmb{1}  +\bigg(\sum_{i=1}^3 \{S^i , G\} \bigg) + O(e^4)\,.
\end{equation}
Now we have
\begin{equation}
    \Sigma^2 = s^\prime (s^\prime +1 )\pmb{1} \,,
\end{equation}
where $ s^\prime = s+\delta$ and $s=\frac{1}{2}$ is the original spin of the electron, $\delta$ is the small deviation matrix. By expanding,
\begin{equation}
\begin{aligned}
\Sigma^2 &= (s+\delta )(s+\delta +1) \pmb{1} \\
&= \bigg(\frac{1}{2}+\delta \bigg)\bigg(\frac{1}2+\delta +1 \bigg)\pmb{1} \\
&= \frac{3}{4}\pmb{1} + 2\delta + \delta^2 \\
&\approx \frac{3}{4}\pmb{1} + 2\delta \,,
\end{aligned}
\end{equation}
where $\delta^2$ is very small and can be dropped. Comparing like-terms in equation \ref{eq:11},  we identify the small spin deviation as, in matrix form,
\begin{equation}
    \delta = \frac{\Delta}{2} \,.
\end{equation}
When the gauge field interaction is turned off, i.e. $A^{\mu}=0$, we will also have $G=0$. Then we can see that this gives us back the original spin value of the electron.

\section{Quantization of the effective Dirac algebra}
So far we have investigated the classical theory for effective Dirac algebra by gauge field interaction. Next we will carry out quantization of the classical theory of effective Dirac algebra . First of all, we know that the quantization of the Dirac field is to impose the equal time anti-commutation relation (ETCR) on the Dirac spinors,  for which the spinor field is turned into field operator,
\begin{equation}
\begin{aligned}
 & \{ \hat{\psi}_a (t, \mathrm{\pmb{x}} ) , \hat{\psi}_b^\dagger (t, \mathrm{\pmb{y}} )\} = \delta^3 (\mathrm{\pmb{x}} -\mathrm{\pmb{y}} ) \delta_{ab} \,, \\
 & \{ \hat{\psi}_a (t, \mathrm{\pmb{x}} ) , \hat{\psi}_b (t, \mathrm{\pmb{y}} )\} = \{ \hat{\psi}_a^\dagger (t, \mathrm{\pmb{x}} ) , \hat{\psi}_b^\dagger (t, \mathrm{\pmb{y}} )\} = 0 \,.
\end{aligned}
\end{equation}
This will be the same when we promote to the effective case. We also know that the quantization of the gauge field reads,
\begin{equation} \label{eq:comm}
\begin{aligned}
  &  [\hat{A}_{\mu} (t, \mathrm{\pmb{x}} ) , \hat{\Pi}_\nu (t, \mathrm{\pmb{y}} )  ] = -i\eta_{\mu\nu}\delta^3 (\mathrm{\pmb{x}} -\mathrm{\pmb{y}} ) \,, \\
    &[\hat{A}_{\mu} (t, \mathrm{\pmb{x}} ) ,\hat{A}_{\nu} (t, \mathrm{\pmb{y}} )] =[\hat{\Pi}_{\mu} (t, \mathrm{\pmb{x}} ) ,\hat{\Pi}_{\nu} (t, \mathrm{\pmb{y}} )] =  0 \,,
\end{aligned}
\end{equation}
where  $\hat{\Pi}^\nu = -\hat{\Dot{A}}^\nu$ is the conjugated momentum. This will also be same when we promote to the effective case. However, as in the effective case the curved metric tensor is defined by the gauge fields, see in equations \ref{eq:EffectiveDiracAlgebra} and \ref{eq:metricc}, then the quantization amounts to quantizing the effective Dirac algebra, so is to quantize the metric. This marks the difference from the non-effective case, where in flat space time, the normal Dirac alegbra and the Minkowski metric are not quantized. In other words, the quantized effective Dirac algebra reads, $\{ \hat{\Gamma}^{\mu}_L , \hat{\Gamma}^{\nu}_L \} = 2 \hat{g}^{\mu\nu}\pmb{1}$, which is 

\begin{equation} 
\{ \hat{\Gamma}^{\mu}_L , \hat{\Gamma}^{\nu}_L \} = 2 f^2 [\hat{\slashed{A}}] \eta^{\mu\nu} + \frac{\lambda e}{\pi} f[\hat{\slashed{A}}] \bigg( \frac{1}{1-\frac{\lambda^2 e^2}{4\pi^2}\hat{A}^2} \bigg) ( 2\eta^{\mu\nu}\hat{\slashed{A}}-\gamma^\mu \hat{A}^\nu - \gamma^\nu \hat{A}^\mu )\,.
\end{equation}
And the quantized metric reads,
\begin{equation} 
\hat{g}_{\mu\nu} = \frac{1}{4}\mathrm{Tr}\, \bigg\{ f^2 [\hat{\slashed{A}}] \eta_{\mu\nu} + \frac{\lambda e}{2\pi} f[\hat{\slashed{A}}] \bigg( \frac{1}{1-\frac{\lambda^2 e^2}{4\pi^2}\hat{A}^2} \bigg) ( 2\eta_{\mu\nu}\hat{\slashed{A}}-\gamma_\mu \hat{A}_\nu - \gamma_\nu \hat{A}_\mu )\bigg\} \,.
\end{equation}
Up to second order, by equations(\ref{eq:SecondOrderEffectiveDiracAlgebra}) and (\ref{eq:metriccc}), the quantized versions are 
\begin{equation} 
\{ \hat{\Gamma}^{\mu}_L , \hat{\Gamma}^{\nu}_L \}  = \bigg(2 + \frac{\lambda^2 e^2}{2\pi^2}\hat{A}^2 \bigg)\eta^{\mu\nu}\pmb{1} - \bigg( \frac{\lambda e}{\pi} - \frac{\lambda^2 e^2}{2 \pi^2} \hat{\slashed{A}}   \bigg) (\gamma^{\mu}\hat{A}^\nu + \gamma^{\nu}\hat{A}^\mu ) + O(e^3) \,,
\end{equation}
and
\begin{equation} \label{eq:quantizedMetric}
\hat{g}_{\mu\nu} = \bigg(1 + \frac{\lambda^2 e^2}{4\pi^2}\hat{A}^2 \bigg)\eta_{\mu\nu} +  \frac{\lambda^2 e^2}{2\pi^2}\hat{A}_{\mu} \hat{A}_{\nu} \,.
\end{equation}
It is noticed that explicitly the quantized gauge field is given by
\begin{equation}
\begin{aligned}
\hat{A}_{\mu}(x) &= \int \frac{d^3 \mathrm{\pmb{p}}}{\sqrt{2E_{ \mathrm{\pmb{p}}}}} \sum_{r=0}^3 \Big( a^r_{\mathrm{\pmb{p}}} \epsilon^r_{\mu}(\mathrm{\pmb{p}})e^{-ip\cdot x} + a^{r\dagger}_{\mathrm{\pmb{p}}} \epsilon^r_{\mu}(\mathrm{\pmb{p}}) e^{ip\cdot x} \Big) \,, \\
\end{aligned}
\end{equation}
in which we have to some over all the polarizations of the photon. Now we want to find out the commutation relation of the metric in terms of the quantized gauge fields for the perturbative case. We would like to compute $[\hat{g}_{\mu\nu}(t, \mathrm{\pmb{x}}),\hat{\dot{g}}_{\rho\sigma}(t, \mathrm{\pmb{y}})  ]$, where $\hat{\pi}_{\rho\sigma} =\hat{\dot{g}}_{\rho\sigma}$ is the conjugated momentum of the metric. Differentiating equation (\ref{eq:quantizedMetric}) with respect to time, we obtain
\begin{equation}
\begin{aligned}
    \hat{\dot{g}}_{\mu\nu} &= \frac{\lambda^2 e^2}{4\pi^2}\eta_{\mu\nu}\eta^{\alpha\beta}( \hat{\dot{A}}_{\alpha} \hat{A}_{\beta} +\hat{A}_{\alpha} \hat{\dot{A}}_{\beta}) + \frac{\lambda^2 e^2}{2\pi^2} (\hat{\dot{A}}_{\mu} \hat{A}_{\nu} +\hat{A}_{\mu} \hat{\dot{A}}_{\nu}) \\
    &=-\frac{\lambda^2 e^2}{4\pi^2}\eta_{\mu\nu}\eta^{\alpha\beta}( \hat{\Pi}_{\alpha} \hat{A}_{\beta} +\hat{A}_{\alpha} \hat{\Pi}_{\beta})- \frac{\lambda^2 e^2}{2\pi^2} (\hat{\Pi}_{\mu} \hat{A}_{\nu} +\hat{A}_{\mu} \hat{\Pi}_{\nu}) \,. 
\end{aligned}
\end{equation}
Therefore we have 
\begin{equation}
\begin{aligned}
 &\quad  [\hat{g}_{\mu\nu}(t, \mathrm{\pmb{x}}),\hat{\pi}_{\rho\sigma}(t, \mathrm{\pmb{y}})]\\
 &=\bigg[ \bigg(1 + \frac{\lambda^2 e^2}{4\pi^2}\eta^{\alpha\beta}\hat{A}_{\alpha}(x)\hat{A}_{\beta}(x)  \bigg)\eta_{\mu\nu} + \frac{\lambda^2 e^2}{2\pi^2} \hat{A}_{\mu}(x)\hat{A}_{\nu}(x) \,\,,\,\,\\
 &\quad\quad\quad-\frac{\lambda^2 e^2}{4\pi^2}\eta_{\rho\sigma}\eta^{\lambda\delta}( \hat{\Pi}_{\lambda}(y) \hat{A}_{\delta}(y) +\hat{A}_{\lambda}(y) \hat{\Pi}_{\delta}(y))- \frac{\lambda^2 e^2}{2\pi^2} (\hat{\Pi}_{\rho}(y) \hat{A}_{\sigma}(y) +\hat{A}_{\rho}(y) \hat{\Pi}_{\sigma}(y)) 
 \bigg] \\
 &= \bigg[\frac{\lambda^2 e^2}{4\pi^2}\eta_{\mu\nu}\eta^{\alpha\beta}\hat{A}_{\alpha}(x)\hat{A}_{\beta}(x)  + \frac{\lambda^2 e^2}{2\pi^2} \hat{A}_{\mu}(x)\hat{A}_{\nu}(x) \,\,,\,\,\\
 &\quad\quad\quad-\frac{\lambda^2 e^2}{4\pi^2}\eta_{\rho\sigma}\eta^{\lambda\delta}( \hat{\Pi}_{\lambda}(y) \hat{A}_{\delta}(y) +\hat{A}_{\lambda}(y) \hat{\Pi}_{\delta}(y))- \frac{\lambda^2 e^2}{2\pi^2} (\hat{\Pi}_{\rho}(y) \hat{A}_{\sigma}(y) +\hat{A}_{\rho}(y) \hat{\Pi}_{\sigma}(y)) \bigg]
\end{aligned}
\end{equation}
Next we use the identity of 
\begin{equation} \label{eq:87}
    [A+B , C+D] = [A,C] + [A,D] + [B,C] + [B,D] \,.
\end{equation}
The first term gives
\begin{equation} \label{eq:stepp1}
    \begin{aligned}
    [A,C] &= \bigg[ \frac{\lambda^2 e^2}{4\pi^2}\eta_{\mu\nu}\eta^{\alpha\beta}\hat{A}_{\alpha}(x)\hat{A}_{\beta}(x)\,\, ,\,\, -\frac{\lambda^2 e^2}{4\pi^2}\eta_{\rho\sigma}\eta^{\lambda\delta}( \hat{\Pi}_{\lambda}(y) \hat{A}_{\delta}(y) +\hat{A}_{\lambda}(y) \hat{\Pi}_{\delta}(y))   \bigg]\\
    &=\frac{\lambda^4 e^4}{16\pi^4}\bigg( \Big[ \eta_{\mu\nu}\eta^{\alpha\beta}\hat{A}_{\alpha}(x)\hat{A}_{\beta}(x) , -\eta_{\rho\sigma}\eta^{\lambda\delta}\hat{\Pi}_{\lambda}(y)\hat{A}_{\delta}(y) \Big] + \\    &\quad\quad\quad\quad\quad\quad\quad\quad\quad\Big[\eta_{\mu\nu}\eta^{\alpha\beta}\hat{A}_{\alpha}(x)\hat{A}_{\beta}(x) ,-\eta_{\rho\sigma}\eta^{\lambda\delta}\hat{A}_{\lambda}(y)\hat{\Pi}_{\delta}(y) \Big]    \bigg) \\
    &=-\frac{\lambda^4 e^4}{16\pi^4}\bigg( \eta_{\mu\nu}\eta^{\alpha\beta}\eta_{\rho\sigma}\eta^{\lambda\delta} \Big[ \hat{A}_{\alpha}(x)\hat{A}_\beta (x) , \hat{\Pi}_{\lambda}(y) \hat{A}_{\delta}(y) \Big] \\
    &\quad\quad\quad\quad\quad\quad + \eta_{\mu\nu}\eta^{\alpha\beta}\eta_{\rho\sigma}\eta^{\lambda\delta} \Big[ \hat{A}_{\alpha}(x)\hat{A}_\beta (x) , \hat{A}_{\lambda}(y) \hat{\Pi}_{\delta}(y) \Big] \bigg) \,.
    \end{aligned}
\end{equation}
To proceed, we need to use the identity of
\begin{equation} \label{eq:89}
    [AB,CD] = A[B,C]D + [A,C]BD + CA[B,D] + C[A,D]B .
\end{equation}
Thus the commutators in equation (\ref{eq:stepp1}) are evaluated as follow,
\begin{equation} \label{eq:stepp2}
\begin{aligned}
 &\quad\Big[ \hat{A}_{\alpha}(x)\hat{A}_{\beta} (x) , \hat{\Pi}_{\lambda}(y) \hat{A}_{\delta}(y) \Big]\\
 &=\hat{A}_{\alpha}(x)[\hat{A}_{\beta}(x), \hat{\Pi}_{\lambda}(y)]\hat{A}_{\delta}(y) + [\hat{A}_{\alpha}(x) , \hat{\Pi}_{\lambda}(y) ]\hat{A}_{\beta}(x) \hat{A}_{\delta}(y) \\
 &\quad\quad\quad + \hat{\Pi}_{\lambda}(y) \hat{A}_{\alpha}(x) [\hat{A}_{\beta} (x) , \hat{A}_{\delta}(y) ] + \hat{\Pi}_{\lambda}(y) [\hat{A}_{\alpha}(x) ,  \hat{A}_{\delta}(y) ] \hat{A}_{\beta}(x) \\
 &=-i\eta_{\beta\lambda}\delta^3 (\mathrm{\pmb{x}}-\mathrm{\pmb{y}}) \hat{A}_{\alpha}(x)\hat{A}_{\delta}(y)-i\eta_{\alpha\lambda}\hat{A}_{\beta}(x)\hat{A}_{\delta}(y) \,,
\end{aligned}
\end{equation}
where in the last line we have used the commutation relation in equation (\ref{eq:comm}). Next we have
\begin{equation}  \label{eq:stepp3}
\begin{aligned} 
&\quad\Big[ \hat{A}_{\alpha}(x)\hat{A}_\beta (x), \hat{A}_{\lambda}(y) \hat{\Pi}_{\delta}(y)
\Big]\\
&=  \hat{A}_{\alpha}(x)[\hat{A}_\beta (x), \hat{A}_{\lambda}(y)   ]\hat{\Pi}_{\delta}(y) + [\hat{A}_{\alpha}(x) , \hat{A}_{\lambda}(y) ]\hat{A}_\beta (x)\hat{\Pi}_{\delta}(y) \\
&\quad\quad\quad+ \hat{A}_{\lambda}(y)\hat{A}_{\alpha}(x)[\hat{A}_\beta (x), \hat{\Pi}_{\delta}(y)] + \hat{A}_{\lambda}(y)[\hat{A}_{\alpha}(x) ,\hat{\Pi}_{\delta}(y) ]\hat{A}_\beta (x) \\
&= -i\eta_{\beta\delta}\delta^3 (\mathrm{\pmb{x}}-\mathrm{\pmb{y}}  )\hat{A}_{\lambda}(y)\hat{A}_{\alpha}(x) -i\eta_{\alpha\delta}\delta^3 (\mathrm{\pmb{x}}-\mathrm{\pmb{y}})\hat{A}_{\lambda}(y)\hat{A}_\beta (x) \,.
\end{aligned}
\end{equation}
Putting the results of (\ref{eq:stepp2}) and (\ref{eq:stepp3}) back into (\ref{eq:stepp1}), we obtain
\begin{equation} \label{eq:block1}
 \begin{aligned}
[A,C] &= \frac{i\lambda^4 e^4}{16\pi^4}\delta^3 (\mathrm{\pmb{x}}-\mathrm{\pmb{y}}) \bigg( \eta_{\mu\nu}\eta^{\alpha\delta}\eta_{\rho\sigma} \hat{A}_{\alpha}(x)\hat{A}_{\delta}(y) +\eta_{\mu\nu}\eta_{\rho\sigma}\eta^{\beta\delta}\hat{A}_{\beta}(x)\hat{A}_{\delta}(y) \\
&\quad\quad\quad\quad + \eta_{\mu\nu}\eta_{\rho\sigma}\eta^{\lambda\alpha}\hat{A}_{\lambda}(y)\hat{A}_{\alpha}(x) +  \eta_{\mu\nu}\eta_{\rho\sigma}\eta^{\lambda\beta}\hat{A}_{\lambda}(y)\hat{A}_{\beta}(x)   \bigg) \\
&= \frac{i\lambda^4 e^4}{4\pi^4}\delta^3 (\mathrm{\pmb{x}}-\mathrm{\pmb{y}}) \eta_{\mu\nu}\eta_{\rho\sigma}\hat{A}_{\alpha}(x)\hat{A}^{\alpha}(y) \,.
 \end{aligned}
\end{equation}
Next we have
\begin{equation} \label{eq:block2}
\begin{aligned}
[A,D]  &=  \bigg[ \frac{\lambda^2 e^2}{4\pi^2}\eta_{\mu\nu}\eta^{\alpha\beta}\hat{A}_{\alpha}(x)\hat{A}_{\beta}(x) , - \frac{\lambda^2 e^2}{2\pi^2} (\hat{\Pi}_{\rho}(y) \hat{A}_{\sigma}(y) +\hat{A}_{\rho}(y) \hat{\Pi}_{\sigma}(y))  \bigg] \\
&=-\frac{\lambda^4 e^4}{8\pi^4}\bigg(\eta_{\mu\nu}\eta^{\alpha\beta}[\hat{A}_{\alpha}(x)\hat{A}_\beta (x), \hat{\Pi}_{\rho}(y)\hat{A}_{\sigma}(y)] +  \eta_{\mu\nu}\eta^{\alpha\beta}[\hat{A}_{\alpha}(x)\hat{A}_\beta (x), \hat{A}_{\rho}(y)\hat{\Pi}_{\sigma}(y) ]    \bigg) \\
&=-\frac{\lambda^4 e^4}{8\pi^4}\bigg(\eta_{\mu\nu}\eta^{\alpha\beta}(-i\eta_{\beta\rho}\delta^3 (\mathrm{\pmb{x}}-\mathrm{\pmb{y}})\hat{A}_{\alpha}(x)\hat{A}_{\sigma}(y) -i\eta_{\alpha\rho}\delta^3 (\mathrm{\pmb{x}}-\mathrm{\pmb{y}})\hat{A}_{\beta}(x)\hat{A}_{\sigma}(y) )\\
&\quad\quad\quad + \eta_{\mu\nu}\eta^{\alpha\beta}( -i\eta_{\alpha\sigma}\delta^3 (\mathrm{\pmb{x}}-\mathrm{\pmb{y}})\hat{A}_{\rho}(y)\hat{A}_{\beta}(x) -i\eta_{\beta\sigma} \delta^3 (\mathrm{\pmb{x}}-\mathrm{\pmb{y}})\hat{A}_{\rho}(y)\hat{A}_{\alpha}(x) ) \bigg) \\
&=\frac{i\lambda^4 e^4}{4\pi^4}\delta^3 (\mathrm{\pmb{x}}-\mathrm{\pmb{y}}) \eta_{\mu\nu}(\hat{A}_{\rho}(x)\hat{A}_{\sigma}(y) +  \hat{A}_{\rho}(y)\hat{A}_{\sigma}(x)) \,.
\end{aligned}
\end{equation}
Next we have, 
\begin{equation}      \label{eq:block3}
\begin{aligned}
[B,C] &= \bigg[ \frac{\lambda^2 e^2}{2\pi^2} \hat{A}_{\mu}(x)\hat{A}_{\nu}(x) , -\frac{\lambda^2 e^2}{4\pi^2}\eta_{\rho\sigma}\eta^{\lambda\delta}( \hat{\Pi}_{\lambda}(y) \hat{A}_{\delta}(y) +\hat{A}_{\lambda}(y) \hat{\Pi}_{\delta}(y)) \bigg] \\
&=-\frac{\lambda^4 e^4}{8\pi^4}\bigg(\eta_{\rho\sigma}\eta^{\lambda\delta}[\hat{A}_{\mu}(x)\hat{A}_{\nu}(x), \hat{\Pi}_{\lambda}(y)\hat{A}_{\delta}(y)] + \eta_{\rho\sigma}\eta^{\lambda\delta}[\hat{A}_{\mu}(x)\hat{A}_{\nu}(x), \hat{A}_{\lambda}(y)\hat{\Pi}_{\delta}(y)        ] \bigg) \\
&=-\frac{\lambda^4 e^4}{8\pi^4}\bigg( \eta_{\rho\sigma}\eta^{\lambda\delta}( -i\eta_{\nu\lambda}\delta^3 (\mathrm{\pmb{x}}-\mathrm{\pmb{y}})\hat{A}_{\mu}(x)\hat{A}_{\delta}(y) -i\eta_{\mu\lambda} \delta^3 (\mathrm{\pmb{x}}-\mathrm{\pmb{y}}) \hat{A}_{\nu}(x)\hat{A}_{\delta}(y))   \\
& \quad\quad\quad + \eta_{\rho\sigma}\eta^{\lambda\delta}(-i\eta_{\mu\delta}\delta^3 (\mathrm{\pmb{x}}-\mathrm{\pmb{y}})\hat{A}_{\lambda}(y)\hat{A}_{\nu}(x) -i\eta_{\nu\delta}\delta^3 (\mathrm{\pmb{x}}-\mathrm{\pmb{y}})\hat{A}_{\lambda}(y)\hat{A}_{\mu}(x))   \bigg) \\
&=\frac{i\lambda^4 e^4}{4\pi^4}\delta^3 (\mathrm{\pmb{x}}-\mathrm{\pmb{y}})\eta_{\rho\sigma}(\hat{A}_{\mu}(x) \hat{A}_{\nu}(y) +\hat{A}_{\mu}(y) \hat{A}_{\nu}(x)) \,.
\end{aligned}
\end{equation}
Finally we have,
\begin{equation}  \label{eq:block4}
\begin{aligned}
[B,D] &= \bigg[\frac{\lambda^2 e^2}{2\pi^2} \hat{A}_{\mu}(x)\hat{A}_{\nu}(x) , - \frac{\lambda^2 e^2}{2\pi^2} (\hat{\Pi}_{\rho}(y) \hat{A}_{\sigma}(y) +\hat{A}_{\rho}(y) \hat{\Pi}_{\sigma}(y))
\bigg] \\
&= -\frac{\lambda^4 e^4}{4\pi^4} \bigg( [\hat{A}_{\mu}(x)\hat{A}_{\nu}(x) , \hat{\Pi}_{\rho}(y)\hat{A}_{\sigma}(y)] +  [\hat{A}_{\mu}(x)\hat{A}_{\nu}(x) ,\hat{A}_{\rho}(y)\hat{\Pi}_{\sigma}(y) ] \bigg) \\
&=-\frac{\lambda^4 e^4}{4\pi^4} \bigg(-i\eta_{\nu\rho}\delta^3 (\mathrm{\pmb{x}}-\mathrm{\pmb{y}})\hat{A}_{\mu}(x)\hat{A}_{\sigma}(y)-i\eta_{\mu\rho}\delta^3 (\mathrm{\pmb{x}}-\mathrm{\pmb{y}})\hat{A}_{\nu}(x)\hat{A}_{\sigma}(y) \\
&\quad\quad\quad -i\eta_{\nu\sigma}\delta^3 (\mathrm{\pmb{x}}-\mathrm{\pmb{y}})\hat{A}_{\mu}(x)\hat{A}_{\rho}(y) -i\eta_{\mu\sigma}\delta^3 (\mathrm{\pmb{x}}-\mathrm{\pmb{y}})\hat{A}_{\nu}(x)\hat{A}_{\rho}(y) \bigg) \\
&=\frac{i\lambda^4 e^4}{4\pi^4} \delta^3 (\mathrm{\pmb{x}}-\mathrm{\pmb{y}})(\eta_{\nu\rho} \hat{A}_{\mu}(x) \hat{A}_{\sigma}(y) +\eta_{\mu\rho}\hat{A}_{\nu}(x)\hat{A}_{\sigma}(y) + \eta_{\nu\sigma}\hat{A}_{\mu}(x)\hat{A}_{\rho}(y) +\eta_{\mu\sigma}\hat{A}_{\nu}(x)\hat{A}_{\rho}(y)  ) \,.
\end{aligned}
\end{equation}
Therefore, combining all the results of (\ref{eq:block1}), (\ref{eq:block2}), (\ref{eq:block3}) and (\ref{eq:block4}), we obtain the canonical quantization of the metric in terms of quantized gauge field,
\begin{equation}
\begin{aligned}
&\quad[\hat{g}_{\mu\nu}(t, \mathrm{\pmb{x}}),\hat{\pi}_{\rho\sigma}(t, \mathrm{\pmb{y}})  ] \\ &=\frac{i\lambda^4 e^4}{4\pi^4} \delta^3 (\mathrm{\pmb{x}}-\mathrm{\pmb{y}}) \Big( \eta_{\mu\nu}\eta_{\rho\sigma}\hat{A}_{\alpha}(x)\hat{A}^{\alpha}(y)+ \eta_{\mu\nu}(\hat{A}_{\rho}(x)\hat{A}_{\sigma}(y) +  \hat{A}_{\rho}(y)\hat{A}_{\sigma}(x)) \\
& \quad\quad+  \eta_{\rho\sigma}(\hat{A}_{\mu}(x) \hat{A}_{\nu}(y) +\hat{A}_{\mu}(y) \hat{A}_{\nu}(x)) + \eta_{\nu\rho} \hat{A}_{\mu}(x) \hat{A}_{\sigma}(y) +\eta_{\mu\rho}\hat{A}_{\nu}(x)\hat{A}_{\sigma}(y) \\
&\quad\quad+\eta_{\nu\sigma}\hat{A}_{\mu}(x)\hat{A}_{\rho}(y) +\eta_{\mu\sigma}\hat{A}_{\nu}(x)\hat{A}_{\rho}(y)  \Big) \,.
\end{aligned}
\end{equation}
It has to be remarked that $x^0 =y^0$ for the gauge field operators for which the equal-time commutation has been imposed. So it is meant to be $\hat{A}_{\mu}(x) = \hat{A}_{\mu}(t, \mathrm{\pmb{x}})  $ and $\hat{A}_{\nu}(y) = \hat{A}_{\nu}(t, \mathrm{\pmb{y}})  $, and we just write $\hat{A}_{\mu}(x)$ and $\hat{A}_{\mu}(y)$ to save the space.
Next we have 
\begin{equation}
    [\hat{g}_{\mu\nu}(t, \mathrm{\pmb{x}}),\hat{g}_{\rho\sigma}(t, \mathrm{\pmb{y}})] =0 \,.
\end{equation}
This is very easy to show. This because, using the identity \ref{eq:87} and \ref{eq:89}, it can be easily seen that after the expansion, we will have all terms that involve the commutator of the gauge field operators $[\hat{A}_{\mu}(x), \hat{A}_{\nu}(y)   ]  $, which is zero by equation \ref{eq:comm}.  It follows that $[\hat{g}_{\mu\nu}(t, \mathrm{\pmb{x}}),\hat{g}_{\rho\sigma}(t, \mathrm{\pmb{y}})] =0$.

\section{Application to the relativistic hydrogen atom}
In this section, we would apply the results we have to an actual electromagnetic system. For example, here we would choose to study the relativistic hydrogen atom, and see how such a quantum dynamic system redefines the metric of spacetime when the gauge field interaction is turned on. 

First, we are interested in the system of a relativistic electron with charge $e$ circulating around the proton with charge $+e$. Therefore, we take the gauge field vector as 
\begin{equation}
    A^{\mu} = 
    \begin{pmatrix}
     -\frac{Ze}{4\pi \epsilon_0 r}\\
     0 \\
     0\\
     0\\
    \end{pmatrix} \,,
\end{equation}
where $A^0 = V = -\frac{Ze}{4\pi\epsilon_0 r}$ is the central potential and $A^i =0$, and $Z$ is the atomic number for which $Z=1$ for our hydrogen case. We will use the metric convention as diag(1,-1,-1,-1), so $A_0 = A^0$ and $A_i = -A^i$. 

Recalling the Dirac equation in \ref{eq:DiracEq}, now with the hydrogen potential then we obtain the Dirac equation for the relativistic hydrogen atom with central potential.
\begin{equation}
i\hbar \gamma^0 \bigg( \frac{1}{c}\frac{\partial}{\partial t} -i\frac{Z e^2}{4\pi\epsilon_0 \hbar cr} \bigg) \psi + i\hbar\gamma^i\partial_i \psi -mc\psi = 0 \,.
\end{equation}

Now recalling the full metric result we had in equation \ref{eq:metricc}, we need to put back the information of $A_0$ and then compute the matrices and its trace. First of all, notice that
\begin{equation}
    \slashed{A}= A_{\mu}\gamma^\mu = A_0 \gamma^0 + A_i \gamma^i = A_0 \gamma^0 \,,
\end{equation}
as all $A_i$s are zero. Also notice that
\begin{equation}
    A^2 = A_{\mu}A^{\mu} = A_0 A^0 + A_i A^i = (A_0 )^2 = \bigg(\frac{Ze}{4\pi\epsilon_0 r}\bigg)^2 \,.
\end{equation}
And the functional in (\ref{eq:functional}) now becomes
\begin{equation}
    f[\slashed{A}] = \frac{1}{1+\frac{\lambda e} {2\pi}A_0 \gamma^0 } = \Big(1 + \frac{\lambda e} {2\pi}A_0 \gamma^0 \Big)^{-1} \,.
\end{equation}
Then now we write the equation (\ref{eq:metricc}) as 
\begin{equation}
\begin{aligned}
g_{\mu\nu} &=\frac{1}{4}\mathrm{Tr}\Bigg( \bigg(\big(1 + \frac{\lambda e} {2\pi}A_0 \gamma^0 \big)^{-1}\bigg)^2  \eta_{\mu\nu} + \frac{\lambda e}{2\pi}\big(1 + \frac{\lambda e} {2\pi}A_0 \gamma^0 \big)^{-1} \bigg(\frac{1}{1-\frac{\lambda^2 e^2}{4\pi^2}A_0^2}\bigg)  \\
&\quad\quad\quad\quad  \quad\quad  \quad\quad\quad\quad  \quad\quad \quad\quad\quad\quad  \quad\quad     \times (2\eta_{\mu\nu} A_0 \gamma^0 - \gamma_{\mu}A_{\nu} - \gamma_\nu \gamma_\mu ) \Bigg) \,.
\end{aligned}
\end{equation}
Next we can carry out analysis on the components of the metric tensor. We have the following.
\begin{equation} \label{eq:104s}
    g_{00} = \frac{1}{4}\mathrm{Tr}\bigg(\big(1 + \frac{\lambda e} {2\pi}A_0 \gamma^0 \big)^{-1}\bigg)^2 \eta_{00}  \,,
\end{equation}
\begin{equation}  \label{eq:105s}
g_{ii}    = \frac{1}{4}\mathrm{Tr}\left(\bigg( \big(1 + \frac{\lambda e} {2\pi}A_0 \gamma^0 \big)^{-1}\bigg)^2 \eta_{ii} +\frac{\lambda e}{2\pi} (1 + \frac{\lambda e} {2\pi}A_0 \gamma^0 \big)^{-1}\bigg(\frac{1}{1-\frac{\lambda^2 e^2}{4\pi^2}A_0^2}\bigg) 2\eta_{ii} A_0 \gamma^0 \right)\,,
\end{equation}
\begin{equation}  \label{eq:106s}
g_{0i} = \frac{1}{4}\mathrm{Tr}\bigg(\frac{\lambda e}{2\pi} (1 + \frac{\lambda e} {2\pi}A_0 \gamma^0 \big)^{-1}  \bigg(\frac{1}{1-\frac{\lambda^2 e^2}{4\pi^2}A_0^2}\bigg) (-\gamma_i A_0  )   \bigg) = g_{i0}\,,
\end{equation}
\begin{equation}  \label{eq:107s}
  g_{ij} = g_{ji} =0 \quad \quad\text{for}\,\, i \neq j \,.  
\end{equation}
To explicitly compute the metric tensor, we need to use a specific representation of the Dirac matrices. For convenience, we will pick the Dirac representation, which are as follow,
\begin{equation}
    \gamma^0 =
    \begin{pmatrix}
    1 & 0 & 0 & 0 \\
    0 & 1 & 0 & 0 \\
    0 & 0 & -1 & 0 \\
    0 & 0 & 0 & -1
    \end{pmatrix}\,\,,\,\,
    \gamma^1 =
    \begin{pmatrix}
    0 & 0 & 0 & 1 \\
    0 & 0 & 1 & 0 \\
    0 & -1 & 0 & 0 \\
    -1 & 0 & 0 & 0
    \end{pmatrix}\,\,,\,\,
    \gamma^2 =
    \begin{pmatrix}
    0 & 0 & 0 & -i \\
    0 & 0 & i & 0 \\
    0 & i & 0 & 0 \\
    -i & 0 & 0 & 0
    \end{pmatrix}\,\,,\,\,
    \gamma^3 =
    \begin{pmatrix}
    0 & 0 & 1 & 0 \\
    0 & 0 & 0 & -1 \\
    -1 & 0 & 0 & 0 \\
    0 & 1 & 0 & 0
    \end{pmatrix}\,\,.\,\,
\end{equation}
Therefore, now we can explicitly compute the metric tensor components. First we evaluate the inverse matrix,
\begin{equation}
\begin{aligned}
    (1 + \frac{\lambda e} {2\pi}A_0 \gamma^0 \big)^{-1} = 
    \begin{pmatrix}
    1+\frac{\lambda e}{2\pi}A_0 &           0              & 0  &    0                    \\
            0                &1+\frac{\lambda e}{2\pi}A_0  & 0   &    0                       \\
            0                &           0              & 1-\frac{\lambda e}{2\pi}A_0 &0      \\
            0                &           0              &   0 & 1-\frac{\lambda e}{2\pi}A_0
    \end{pmatrix}^{-1} \,.
\end{aligned}
\end{equation}
For simplicity, define $a=\frac{\lambda e}{2\pi}A_0$. And since $\eta_{00}=1$, then by equation \ref{eq:104s}
\begin{equation} \label{eq:110s}
\begin{aligned}
    g_{00}&=\frac{1}{4}\mathrm{Tr}\bigg(\big(1 + \frac{\lambda e} {2\pi}A_0 \gamma^0 \big)^{-1}\bigg)^2 \eta_{00} \\  
    &= \frac{1}{4}\mathrm{Tr}
    \begin{pmatrix}
    \frac{1}{1+a} & 0 & 0 & 0\\
    0             &\frac{1}{1+a} &0 & 0 \\
    0 & 0 & \frac{1}{1-a} & 0 \\
    0 & 0 & 0 &\frac{1}{1-a}
    \end{pmatrix}^2 \\
    &=\frac{1}{2}\bigg( \frac{1}{1+a} \bigg)^2 + \frac{1}{2}\bigg( \frac{1}{1-a} \bigg)^2 \\
    &=\frac{1}{2}\bigg( \frac{1}{1+\frac{\lambda e}{2\pi}A_0} \bigg)^2 + \frac{1}{2}\bigg( \frac{1}{1-\frac{\lambda e}{2\pi}A_0} \bigg)^2 \,.
    \end{aligned}
\end{equation}
Next we need to find $g_{ii}$. As $\eta_{ii}= -1$, then by equation \ref{eq:105s} we have
\begin{equation} \label{eq:111s}
\begin{aligned}
g_{ii}&=\frac{1}{4}\mathrm{Tr}
\begin{pmatrix}
-\Big(\frac{1}{1+a}\Big)^2      &       0         &        0      &          0  \\
                 0              & -\Big(\frac{1}{1+a}\Big)^2   & 0    &          0\\
                 0              &       0         &   -\Big(\frac{1}{1-a}\Big)^2 & 0\\
0 & 0 & 0 & -\Big(\frac{1}{1-a}\Big)^2 
\end{pmatrix}    \\
&\quad\quad\quad\quad+\frac{1}{4}\bigg(\frac{-2a}{1-a^2}\bigg) \mathrm{Tr} \left(
\begin{pmatrix}
    \frac{1}{1+a} & 0 & 0 & 0\\
    0             &\frac{1}{1+a} &0 & 0 \\
    0 & 0 & \frac{1}{1-a} & 0 \\
    0 & 0 & 0 &\frac{1}{1-a}
\end{pmatrix}
\begin{pmatrix}
1 & 0 & 0 & 0 \\
    0 & 1 & 0 & 0 \\
    0 & 0 & -1 & 0 \\
    0 & 0 & 0 & -1
\end{pmatrix} \right) \\
&= -\frac{1}{2} \bigg( \Big(\frac{1}{1+a}\Big)^2 +   \Big(\frac{1}{1-a}\Big)^2  \bigg) -\frac{a}{1-a^2}\bigg(\frac{1}{1+a} - \frac{1}{1-a} \bigg) \\
&=  -\frac{1}{2} \bigg( \Big(\frac{1}{1+\frac{\lambda e}{2\pi}A_0}\Big)^2 +   \Big(\frac{1}{1-\frac{\lambda e}{2\pi}A_0}\Big)^2  \bigg) -\frac{\frac{\lambda e}{2\pi}A_0}{1-\Big(\frac{\lambda e}{2\pi}A_0\Big)^2}\bigg(\frac{1}{1+\frac{\lambda e}{2\pi}A_0} - \frac{1}{1-\frac{\lambda e}{2\pi}A_0} \bigg) \,.
\end{aligned}
\end{equation}
Finally, we compute $g_{0i}$. First consider $g_{01}$, using equation \ref{eq:106s} we get
\begin{equation}
    g_{01} = -\frac{1}{4} \bigg(\frac{a}{1-a^2}  \bigg)\mathrm{Tr}\left( \begin{pmatrix}
    \frac{1}{1+a} & 0 & 0 & 0\\
    0             &\frac{1}{1+a} &0 & 0 \\
    0 & 0 & \frac{1}{1-a} & 0 \\
    0 & 0 & 0 &\frac{1}{1-a}
\end{pmatrix}
    \begin{pmatrix}
0 & 0 & 0 & 1\\
    0 & 0 & 1 & 0 \\
    0 & -1 & 0 & 0 \\
    -1 & 0 & 0 & 0
\end{pmatrix}
       \right) =0 \,.
\end{equation}
Then similarly we find $g_{02}$ is 
\begin{equation}
    g_{01} = -\frac{1}{4} \bigg(\frac{a}{1-a^2}  \bigg)\mathrm{Tr}\left( \begin{pmatrix}
    \frac{1}{1+a} & 0 & 0 & 0\\
    0             &\frac{1}{1+a} &0 & 0 \\
    0 & 0 & \frac{1}{1-a} & 0 \\
    0 & 0 & 0 &\frac{1}{1-a}
\end{pmatrix}
    \begin{pmatrix}
0 & 0 & 0 & -i\\
    0 & 0 & i & 0 \\
    0 & i & 0 & 0 \\
    -i & 0 & 0 & 0
\end{pmatrix}
       \right) =0 \,.
\end{equation}
Then finally we find $g_{03}$ is
\begin{equation}
    g_{03} = -\frac{1}{4} \bigg(\frac{a}{1-a^2}  \bigg)\mathrm{Tr}\left( \begin{pmatrix}
    \frac{1}{1+a} & 0 & 0 & 0\\
    0             &\frac{1}{1+a} &0 & 0 \\
    0 & 0 & \frac{1}{1-a} & 0 \\
    0 & 0 & 0 &\frac{1}{1-a}
\end{pmatrix}
    \begin{pmatrix}
    0 & 0 & 1 & 0\\
    0 & 0 & 0 & -1 \\
    -1 & 0 & 0 & 0 \\
    0 & 1 & 0 & 0
\end{pmatrix}
       \right) =0 \,.
\end{equation}
Therefore, we can see that the metric tensor contains diagonal terms only. Also we can cross check that, when $A_{\mu} $ is turned off, i.e. $A_0 =0$ (or $r\rightarrow\infty$ in our case), by equation (\ref{eq:110s}) we see that $g_{00}=1$ and by equation (\ref{eq:111s})we see that $g_{ii}=-1$. In other words, when the electromagnetic potential is turned off, we obtain back the original Minkowski metric as expected. 

Hence, now we can write down the full metric in spherical coordinates as follow,
\begin{equation} \label{eq:fullmetric}
\begin{aligned}
   & ds^2 = \bigg[   \frac{1}{2}\bigg( \frac{1}{1-\frac{Ze^2 h}{4\pi^2 m c \epsilon_0 r}} \bigg)^2 + \frac{1}{2}\bigg( \frac{1}{1+\frac{Ze^2 h}{4\pi^2 m c \epsilon_0 r}} \bigg)^2 \bigg] dt^2 \\
   &- \bigg[\frac{1}{2} \bigg( \Big(\frac{1}{1-\frac{Ze^2 h}{4\pi^2 m c \epsilon_0 r}}\Big)^2 +   \Big(\frac{1}{1+\frac{Ze^2 h}{4\pi^2 m c \epsilon_0 r}}\Big)^2  \bigg) +\frac{\frac{Ze^2 h}{4\pi^2 m c \epsilon_0 r}}{1-\Big(\frac{Ze^2 h}{4\pi^2 m c \epsilon_0 r}\Big)^2}\bigg(\frac{1}{1-\frac{Ze^2 h}{4\pi^2 m c \epsilon_0 r}} - \frac{1}{1+\frac{Ze^2 h}{4\pi^2 m c \epsilon_0 r}} \bigg) \bigg] \\
   & \times (dr^2 + r^2 d\theta^2 + r^2 \sin^2 \theta d\phi^2 ) \,.
\end{aligned}
\end{equation}
Therefore, the system of relativistic hydrogen atom curves spacetime, given by the metric in \ref{eq:fullmetric}, and imposes gravitational effect.

\section{Discussion on future work for the non-abelian case}
In this discussion section, we would like to give a brief reivew on the difficulty in applying the above results in section 2 to the non-abelian case. We will see that in fact, such generalization is highly non-trivial and we cannot apply the derivation in section 2 due to the non-abelian nature, and we will see why. This demands future work. For the non-abelian case, the Dirac equation with interaction is 
\begin{equation}
i\hbar \gamma^\mu \partial_{\mu}\Psi -g \hbar A_{\mu}  \gamma^\mu \Psi - mc \Psi =0 \,,
\end{equation}
where $A_{\mu} = A_{\mu}^a t^a$ and $t^a$ is the generator of the Lie group, for example SU(3) for coloured gluons. The full Dirac spinor $\Psi$ is the tensor product of the colour state to the normal Dirac spinor,
\begin{equation}
\Psi(x) = | c \rangle \otimes \psi(x) \,,
\end{equation} 
where $| c \rangle $ is the $3 \times 1$ column colour vector that transforms according to the SU(3) group. The $\Psi $ is then a $12 \times 1$ column spinor.  If we explicitly investigate the term
\begin{equation}
A_{\mu}^a t^a \gamma^\mu \,,
\end{equation}
one can see that in general $t^a $ and $\gamma^\mu$ are of different dimensions. For example in the SU(3) case $t^a$ are $3\times3$ Gellmann matrices, while $\gamma^\mu$ are $4\times 4$ matrices. Since by noting that the full spinor $\Psi$ is factorized into colour state and normal Dirac spinor, therefore the term $A_{\mu}^a t^a \gamma^\mu $ means to be  $A_{\mu}^a t^a \otimes \gamma^\mu $. Therefore the Dirac equation for non-abelian interaction means to be
\begin{equation} \label{eq:NonAbelianDiracEquation}
\bigg( i \hbar( \tilde{\pmb{1}} \otimes \gamma^\mu  ) \partial_{\mu} -g \hbar A_\mu \otimes \gamma^\mu - mc \tilde{\pmb{1}} \otimes \pmb{1} \bigg) ( |c \rangle \otimes \psi ) = 0 \,,
\end{equation}
where $\tilde{\pmb{1}}$ is the $3\times3$ identity matrix associated to the SU(3) generator matrices, while $\pmb{1}$ is the $4\times 4$ identity matrix associated to the gamma matrices. Then we can write
\begin{equation} \label{eq:NonAbelianDiractEquation2}
\Psi(x) = \frac{i\hbar}{mc} \left( \frac{1}{\tilde{\pmb{1}}\otimes \pmb{1} + \frac{\hbar g}{mc} A_{\rho} \otimes \gamma^\rho } \right) ( \tilde{\pmb{1}} \otimes \gamma^\mu ) \partial_{\mu}\Psi (x) \,.
\end{equation}  
 
It is also beneficial to express the equations in tensor notations. Notice that $A_{\mu ij} = A^a t^a_{ij}$. Also define
\begin{equation}
\Psi_{jq} = ( \,| c \rangle \otimes \psi  )_{jq} = c_j \psi_q \,.
\end{equation}
Therefore the tensor components of interaction term acting on the full spinor reads
\begin{equation}
\big[(A_{\mu} \otimes \gamma^\mu ) \Psi \big]_{ip} = A_{\mu ij}\gamma^\mu_{pq}\Psi_{jq} \,.
\end{equation}
Then it follows that in tensor notation, equations (\ref{eq:NonAbelianDiracEquation}) is
\begin{equation}
( i\hbar \tilde{\delta}_{ij}\gamma^\mu_{pq} \partial_\mu - g\hbar A_{\mu ij}\gamma^\mu_{pq} - mc \tilde{\delta}_{ij}\delta_{pq} ) \Psi_{jq}(x) = 0 \,.
\end{equation}
And that for equation (\ref{eq:NonAbelianDiractEquation2}) is
\begin{equation}
\Psi_{ip}(x) = \frac{i\lambda}{2\pi} \left( \frac{1}{\tilde{\pmb{1}}\otimes \pmb{1} + \frac{\lambda g}{2\pi} A_{\rho}\otimes \gamma^\rho  } \right)_{ijpq} ( \tilde{\pmb{1}} \otimes \gamma^{\mu}  )_{jaqb} \partial_{\mu}\Psi_{ab}(x)  \,.
\end{equation}
So far so good. However, now consider the functional,
\begin{equation} \label{eq:106}
    f[A_{\rho}\otimes \gamma^{\rho} ] = \frac{1}{\tilde{\pmb{1}}\otimes \pmb{1} + \frac{\lambda g}{2\pi} A_{\rho}\otimes \gamma^\rho } \,.
\end{equation}
We cannot do the expansion analysis as we have in equation (\ref{eq:expansion}) This is because $\slashed{A}\slashed{A}=A^2$ only applies for the abelian case. For non-abelian case, this is not true. We can check the following,
\begin{equation} \label{eq:107}
 \begin{aligned}
 (A_{\rho}\otimes \gamma^{\rho} )(A_{\sigma} \otimes \gamma^{\sigma}) &=  (A_{\rho}^a t^a \otimes \gamma^{\rho} )(A_{\sigma}^b t^b \otimes \gamma^{\sigma}) \\
 &= A_{\rho}^a t^a A_{\sigma}^b t^b \otimes \gamma^{\rho}\gamma^{\sigma} \\
 &= A_{\rho}^a A_{\sigma}^b t^a t^b \otimes \gamma^{\rho}\gamma^{\sigma} \\
 &= \frac{1}{2}( A_{\rho}^a A_{\sigma}^b t^a t^b \otimes \gamma^{\rho}\gamma^{\sigma} + A_{\sigma}^b A_{\rho}^a t^b t^a \otimes \gamma^{\sigma}\gamma^{\rho}) \,.
 \end{aligned}
\end{equation}
But since $A_{\rho}^a A_{\sigma}^b t^a t^b \neq A_{\sigma}^b A_{\rho}^a t^b t^a  $, i.e. $A_{\rho}A_{\sigma} \neq A_{\sigma} A_{\rho}$ due to the fact that the generators do not commute (as $[t^a , t^b] = if^{abc}t^c $), so (\ref{eq:107}) cannot be further factorized. Therefore we conclude that $(A_{\rho}\otimes \gamma^{\rho} )(A_{\sigma} \otimes \gamma^{\sigma}) \neq A^2$. Instead, there is an extra-term correction by the structural constant, we see that from the last line of equation (\ref{eq:107}),
\begin{equation}
    \begin{aligned}
    (A_{\rho}\otimes \gamma^{\rho} )(A_{\sigma} \otimes \gamma^{\sigma}) &= \frac{1}{2}( A_{\rho}^a A_{\sigma}^b t^a t^b \otimes \gamma^{\rho}\gamma^{\sigma} + A_{\sigma}^b A_{\rho}^a (t^a t^b -if^{abc}t^c )\otimes \gamma^{\sigma}\gamma^{\rho}) \\
    &=\frac{1}{2}(A_{\rho}^a A_{\sigma}^b t^a t^b \otimes \gamma^{\rho}\gamma^{\sigma} +  A_{\sigma}^b A_{\rho}^a t^a t^b \otimes \gamma^\sigma \gamma^\rho - if^{abc}t^c A_{\sigma}^b A_{\rho}^a  \otimes \gamma^\sigma \gamma^\rho ) \\
    &= \frac{1}{2}( A_{\rho}^a A_{\sigma}^b t^a t^b \otimes (\gamma^\rho \gamma^\sigma + \gamma^\sigma \gamma^\rho ) -if^{abc}t^c A_{\sigma}^b A_{\rho}^a  \otimes \gamma^\sigma \gamma^\rho ) \\   
    &= \frac{1}{2}( A_{\rho}^a A_{\sigma}^b t^a t^b \otimes (2\eta^{\rho\sigma}\pmb{1} ) -if^{abc}t^c A_{\sigma}^b A_{\rho}^a  \otimes \gamma^\sigma \gamma^\rho ) \\
    &= A^2 \otimes \pmb{1} -\frac{i}{2}f^{abc} A_{\rho}^a A_{\sigma}^b t^c \otimes\gamma^\sigma \gamma^\rho \,.
    \end{aligned}
\end{equation}
Therefore when we carry out series expansion on the functional for the non-abelian case in equation (\ref{eq:106}), the terms will be much more complicated as it involves terms with the structural constant $f^{abc}$ and so on, and this demands rigorous work and a better formalism to be developed in the future. Hence, the generalization of effective Dirac algebra from the abelian case to the non-abelian case is for more complicated than it seems to be.

\section{Conclusion}
In this paper, we have derived classically an explicit formula for the curved metric that is defined by the effective gamma matrix due to gauge field interaction. This shows how the metric can be connected with gauge fields. We also worked out the perturbation metric $h_{\mu\nu}$ by series expansion from the effective gamma matrix and study the geometry contributed by the effective Dirac algebra. Then we demonstrated how the parity operator and spin operator are corrected under the effective gamma matrix in the presence of gauge field interaction, and how the spin of an electron is deviated by a small correction. Then we conduct the canonical quantization of the effective Dirac algebra and hence work out the equal time commutation relation of the metric tensor. Our work shows how to quantize a metric in terms of quantizing gauge fields. Finally, we apply our results to the case of relativistic hydrogen atom and show how such a system results in curving the spacetime metric. This shows how a relativistic, electromagnetic system contribute to changing the geometry of spacetime. At the end we lay some foundations for the future work on the generalization of the effective Dirac algebra to the non-abelian case.


\begin{thebibliography}{}
%
%
\bibitem{ref1}
G. Hooft, and M.J.G. Veltman. One-loop divergencies in the theory of gravitation. \textit{Ann. Inst. H. Poincare Phys. Theor.} A\textbf{20}, 69. 1974.

\bibitem{ref2}
A. Shomer. A pedagogical explanation for the non-renormalizability of gravity. arXiv:0709.3555.

\bibitem{ref3}
S.N.Gupta. Quantization of Einstein’s Gravitational Field: General Treatment. \textit{ Proc.
Phys. Soc. London, Sect. A} \textbf{65}, 608–619. 1952.

\bibitem{ref4}
Z. Bern. Perturbative Quantum Gravity and its Relation to Gauge Theory. \textit{ Living Rev.
Relativity}. \textbf{ 5}. lrr-2002-5. 2002.

\bibitem{ref5}
B.S. DeWitt. Quantum Theory of Gravity. III. Applications of the Covariant Theory.
\textit{Phys. Rev.} \textbf{ 162}. 1239–1256. 1967.

\bibitem{ref6}
B.S. DeWitt. Errata: Quantum Theory of Gravity. \textit{ Phys. Rev.} \textbf{ 171}. 1834. 1968.


\bibitem{ext1}
M. Trzetrzelewski. On the Equivalence Principle and Electrodynamics of Moving Bodies. \textit{EPL (Europhysics Letters)}. Vol \textbf{120}. No. 4. 2018.

\bibitem{ext2}
M. Trzetrzelewski. On the Equivalence Principle and Relativistic Quantum
Mechanics. \textit{Foundation of Physics}. \textbf{50}, p. 1253–1269. 2020.

\bibitem{ext3}
K. Crowther. Effective Spacetime. \textit{Springer}. 2016.

\bibitem{ext4}
J. F. Donoghue. General relativity as an effective field theory: The leading quantum corrections. \textit{Phys.Rev.D} \textbf{50}:3874-3888. 1994

\bibitem{ext5}
J. F. Donoghue. Introduction to the Effective Field Theory Description of Gravity. \textit{Advanced School on Effective Theories: Almunecar, Granada, Spain 26 June-1 July 1995 217–240}. \textit{World Scientific}. 1995.

\bibitem{ext6}
C. P. Burgess. Quantum Gravity in Everyday Life: General Relativity as an Effective Field Theory. \textit{Living Reviews in Relativity.} Vol \textbf{7}. No. 5. 2004. 

\bibitem{ext7}
E. Knox. Effective spacetime geometry. \textit{ Studies in History and Philosophy of Science Part B: Studies in History and Philosophy of Modern Physics}. \textit{Elsevier}. Vol \textbf{44}. Issue 3, p. 346-356. 2013.

\bibitem{ext8}
S.Weinberg. Effective gauge theories. \textit{ Phys. Lett. B}. \textbf{91}. 51–55. 1980.

\bibitem{ext9}
R.L. Arnowitt, S. Deser, and C.W. Misner. Gravitational-Electromagnetic Coupling and
the Classical Self-Energy Problem. \textit{ Phys. Rev.} \textbf{ 120}. 313–320. 1960.

\bibitem{DiracAntiMatter}
Dirac, P. A. M.: A Theory of Electrons and Protons. Proceedings of the Royal Society A. 126 (801): 360–365. (1930)

\bibitem{Peskin}
P. Schroeder and D. V. Schroeder.: An Introduction to Quantum Field Theory. \textit{ABP}, 1995.

\bibitem{Dirac}
Dirac, P. A. M.: The Quantum Theory of the Electron. Proceedings of the Royal Society A: Mathematical, Physical and Engineering Sciences. 117 (778): 610–624. (1928)

\bibitem{Dirac3}
Dirac, P. A. M.: The Lagrangian in quantum mechanics. Phys. Z. Sowjetunion 3, 64–72 (1933)

\bibitem{Julian}
J. Schwinger. On Quantum-Electrodynamics and the Magnetic Moment
of the Electron. \textit{Phys. Rev.} v. 73(4), 416–417. 1948.





\bibitem{Gravitation}
C. W. Misner, K. S. Thorne, J.A. Wheeler.: Gravitation. Princeton University Press. 2017.



\end{thebibliography}
\end{document}